\documentclass[prb,aps,twocolumn,amsmath,amssymb,floatfix,
superscriptaddress]{revtex4}
\usepackage[dvips]{graphics}
\usepackage{color}
\definecolor{dred}{rgb}{0,0,0.6}
\usepackage{graphicx}  
\usepackage{dcolumn}   
\usepackage{bm}        
\usepackage{amssymb}   
\usepackage{amsmath}
\usepackage{braket}
\usepackage{upgreek}
\usepackage[mathscr]{euscript}
\usepackage{color}
\usepackage[toc,page]{appendix}
\usepackage[colorlinks=true, letterpaper=true, pdfstartview=FitV, linkcolor=blue, citecolor=blue, urlcolor=blue]{hyperref}

\begin{document}

\title{First-principles and tight-binding analysis of thermoelectricity in irradiated WSe$_2$}

\author{Cynthia Ihuoma Osuala}
\email{cosuala@stevens.edu}
\affiliation{Department of Physics, Stevens Institute of Technology, Hoboken, NJ 07030, USA}

\author{Tanu Choudhary}
\email{tanunain27@gmail.com}
\affiliation{Department of Physics, Faculty of Natural Sciences, M. S. Ramaiah University of Applied Sciences, Bengaluru 560058, India}

\author{Raju K. Biswas}
\email{rajukumar1718@gmail.com}
\affiliation{Department of Physics, North Eastern Regional Institute of Science and Technology, Nirjuli, Arunachal Pradesh 791109, India}

\author{Sudin Ganguly}
\email{sudinganguly@gmail.com}
\affiliation{Department of Physics, Adamas University, Adamas Knowledge City, Barasat-Barrackpore Road, 24 Parganas North, Kolkata 700126, India}

\author{Santanu K. Maiti}
\email{santanu.maiti@isical.ac.in}
\affiliation{Physics and Applied Mathematics Unit, Indian Statistical
  Institute, 203 Barrackpore Trunk Road, Kolkata 700108, India}

\date{\today}

\begin{abstract}
Electronic and thermoelectric transport in zigzag monolayer WSe$_2$ nanoribbons are studied under monochromatic irradiation. The electronic structure is described within a six-orbital tight-binding framework constructed from the relevant tungsten and selenium orbitals, with atomic spin-orbit coupling included explicitly. Periodic driving is incorporated via the Peierls substitution, and in the high-frequency limit the system is mapped onto an effective static Floquet Hamiltonian with polarization-dependent renormalized hoppings. Coherent transport is evaluated using wave-function matching within the Landauer-B\"{u}ttiker formalism. The lattice thermal conductivity is obtained independently from density functional perturbation theory combined with an iterative solution of the phonon Boltzmann transport equation. Light-induced hopping renormalization reshapes the band dispersion and transmission spectrum near the Fermi level, modifying the Landauer transport integrals that determine electrical and thermal conductances and the Seebeck coefficient. Together with spin-orbit-driven band splitting and reduced lattice thermal conductivity from enhanced anharmonic scattering, this leads to a thermoelectric figure of merit $ZT$ exceeding unity over a broad temperature range.

\end{abstract}

\maketitle
%%%%%%%%%%%%%INTRODUCTION%%%%%%%%%%%%%%%%%%%
\section{\label{sec1} Introduction}
Thermoelectric materials offer a direct route for converting waste heat into usable electrical energy, making them central to technologies ranging from energy harvesting to solid-state refrigeration and waste-heat recovery~\cite{mahan1998,disalvo}. The efficiency of a thermoelectric material is characterized by the dimensionless figure of merit 
ZT, which depends on the Seebeck coefficient, electrical conductivity, and thermal conductivity~\cite{rowe}. Achieving a high value of 
ZT requires a delicate balance between these interdependent quantities, making the design of efficient thermoelectric materials a long-standing challenge~\cite{goldsmid}. As a result, significant research efforts have focused on identifying material platforms and external control strategies capable of enhancing thermoelectric performance~\cite{bell,benenti}.

Low-dimensional materials have emerged as promising candidates for thermoelectric applications due to their enhanced tunability and favorable transport characteristics compared to bulk systems~\cite{hicks1,hicks2,osuala}. In particular, two-dimensional (2D) materials benefit from quantum confinement effects and sharp features in the electronic density of states, which can enhance the Seebeck coefficient by increasing the energy dependence of charge transport near the Fermi level~\cite{mahanpnas,heremans}. Moreover, electronic transmission in nanoscale systems can be efficiently modified by external perturbations such as electric fields, strain, or electromagnetic irradiation, providing alternative routes to optimize thermoelectric response without introducing chemical disorder or structural defects~\cite{etms,benenti}.

Among 2D materials, transition-metal dichalcogenides (TMDs) have attracted considerable attention owing to their intrinsic band gaps, strong spin-orbit coupling (SOC), and rich electronic structures~\cite{butler,geim}. Monolayer tungsten diselenide (WSe$_2$) is particularly appealing, as it exhibits a sizable band gap in the visible range and pronounced SOC-induced band splitting, which strongly influences charge transport near the band edges. These features make WSe$_2$ a promising platform for thermoelectric applications, since the Seebeck coefficient is highly sensitive to band asymmetry and the energy dependence of the electronic transmission function.

Despite these favorable characteristics, controlling and enhancing the thermoelectric response of WSe$_2$ remains challenging. In pristine systems, the electronic transmission function is often approximately symmetric around the Fermi energy, which suppresses the thermopower within the linear response regime~\cite{mahanpnas,etms}. Conventional approaches to breaking this symmetry, such as chemical doping or structural modifications, may introduce disorder and impurity scattering, which can degrade carrier mobility and adversely affect electronic transport~\cite{heremans,li2018}. Therefore, developing alternative and reversible methods for engineering asymmetric electronic transmission is essential for improving thermoelectric performance in WSe$_2$-based devices.

In addition to electronic transport, lattice thermal conductivity plays a crucial role in determining thermoelectric efficiency through its contribution to $ZT$. Spin-orbit coupling is known to influence not only the electronic structure but also phonon properties, particularly in materials containing heavy elements. Recent theoretical studies have demonstrated that SOC can significantly modify phonon anharmonicity and lifetimes, leading to substantial changes in lattice thermal conductivity. For example, Tian et al. showed that the inclusion of SOC in PbSe and PbTe results in longer phonon lifetimes and nearly a twofold enhancement of lattice thermal conductivity at room temperature~\cite{ztian2012}. Similarly, Wu et al. reported that SOC enhances the lattice thermal conductivity of SnSe by up to $\sim 60\%$\cite{wu2019}. In contrast, Li et al. found that SOC has a negligible effect on the lattice thermal conductivity of Mg$_2$Si and Mg$_2$Sn~\cite{wli2012}. These contrasting results indicate that the influence of SOC on phonon transport is highly material dependent and remains an open question.

Monolayer WSe$_2$ contains heavy tungsten atoms and therefore exhibits strong intrinsic SOC, making it an ideal system for investigating SOC-induced effects on lattice thermal transport. While SOC-driven band splitting in WSe$_2$ has been extensively explored in the context of electronic and spin-valley physics~\cite{pyu,dhan}, its role in phonon-mediated heat transport has not yet been fully clarified. Moreover, since intrinsic SOC cannot be experimentally switched off, isolating its contribution to lattice thermal conductivity is challenging. In this context, first-principles calculations provide a controlled and systematic approach to investigate the impact of SOC on phonon transport in WSe$_2$.

Motivated by the need for simultaneous control over electronic and phononic contributions to thermoelectric efficiency, in this work we present a comprehensive study of the thermoelectric properties of monolayer WSe$_2$. We investigate the electronic transport and thermoelectric response under electromagnetic irradiation using a multiorbital tight-binding framework combined with the Landauer-B\"{u}ttiker formalism.  When a material is subjected to irradiation, its electronic hopping parameters can be selectively renormalized through Floquet engineering, leading to anisotropic and direction-dependent modifications of the band structure~\cite{oka2009,kitagawa,lindner}. Such light-induced changes can generate asymmetry in the electronic transmission function, which is a key ingredient for enhancing thermoelectric response\cite{lopez2015,benenti}. Unlike static structural modifications, irradiation provides a flexible and externally tunable means of controlling transport properties without permanently altering the material~\cite{calvo,khoeini}. In parallel, we perform density functional theory-based calculations to study the lattice thermal conductivity of monolayer WSe$_2$ with and without SOC, thereby elucidating the role of SOC in phonon transport. By combining optically tunable electronic transport with first-principles phonon calculations, our work provides a unified understanding of thermoelectric performance in WSe$_2$ and establishes it as a promising platform for tunable thermoelectric applications.

The remainder of this paper is organized as follows. In Section II, we introduce the tight-binding model for WSe$_2$ and outline the theoretical framework used to incorporate electromagnetic irradiation and compute thermoelectric quantities. Section III presents our numerical results and discussion of the light-induced modifications to electronic transport and thermoelectric response. Finally, Section IV summarizes our conclusions.

%%%%%%%%%%%%%%%%%%%%%%%%%%%%%%%%
\section{\label{sec2} Theoretical Model}

In this section, we present the theoretical framework used to calculate the electronic band structure and quantum transport properties of zigzag monolayer WSe$_2$ nanoribbon. Our analysis is based on a six-band tight-binding (TB) model, which has proven to be a powerful and reliable approach for describing the low-energy electronic properties of transition metal dichalcogenides. The formulation follows the model developed by Silva-Guill\'{e}n \textit{et al.}~\cite{silva-guill}.

The six-band real-space Hamiltonian is constructed within the even-symmetry subspace, which consists of three tungsten $d$ orbitals,
$ \{ d_{z^2}, d_{x^2-y^2}, d_{xy} \} $,
and three selenium $p$ orbitals formed from symmetric combinations of the top (t) and bottom (b) chalcogen layers,
\[
\frac{1}{\sqrt{2}}(p_x^t + p_x^b), \quad
\frac{1}{\sqrt{2}}(p_y^t + p_y^b), \quad
\frac{1}{\sqrt{2}}(p_z^t + p_z^b).
\]
Here, the superscripts $t$ and $b$ denote the top and bottom selenium planes, respectively~\cite{silva-guill,khoeini}.

The total Hamiltonian of the system in real space is given by
\begin{equation}
	\begin{aligned}
		H &= \sum_{m,\mu} \epsilon^{\rm W}_{\mu} \,
		a_{m,\mu}^{\dagger} a_{m,\mu}
		+ \sum_{m,\mu} \epsilon^{\rm Se}_{\mu} \,
		b_{m,\mu}^{\dagger} b_{m,\mu} \\
		&\quad + \sum_{m n, \mu \nu}
		\left(
		t^{\text{W-W}}_{m n, \mu \nu} \,
		a_{m,\mu}^{\dagger} a_{n,\nu}
		+ t^{\text{Se-Se}}_{m n, \mu \nu} \,
		b_{m,\mu}^{\dagger} b_{n,\nu}
		\right) \\
		&\quad + \sum_{m n, \mu \nu}
		t^{\text{W-Se}}_{m n, \mu \nu} \,
		a_{m,\mu}^{\dagger} b_{n,\nu}
		+ \text{H.c.}
	\end{aligned}
	\label{eq:1}
\end{equation}

Here, $m$ and $n$ label lattice sites, while $\mu$ and $\nu$ denote atomic orbitals. The operators
$a_{m,\mu}^{\dagger}$ ($b_{m,\mu}^{\dagger}$)
create an electron in orbital $\mu$ at site $m$ on a tungsten (selenium) atom, respectively. The hopping amplitudes $t_{m n, \mu \nu}$ describe electron transfer between orbitals on neighboring sites.

The onsite Hamiltonian for the tungsten $d$ orbitals is expressed as
\begin{equation}
	\epsilon_W =
	\begin{pmatrix}
		\Delta_0 & 0 & 0 \\
		0 & \Delta_2 & - i s \lambda_W \\
		0 & i s \lambda_W & \Delta_2
	\end{pmatrix},
\end{equation}
while the onsite Hamiltonian for the selenium $p$ orbitals is given by
\begin{equation}
	\epsilon_{Se} =
	\begin{pmatrix}
		\Delta_p + V_{pp\pi} & - i s \frac{\lambda_{Se}}{2} & 0 \\
		i s \frac{\lambda_{Se}}{2} & \Delta_p + V_{pp\pi} & 0 \\
		0 & 0 & \Delta_z - V_{pp\pi}
	\end{pmatrix}.
\end{equation}

Here, $s = +1$ ($-1$) corresponds to spin-up (spin-down) states, and $\lambda_W$ and $\lambda_{Se}$ denote the spin-orbit coupling strengths for tungsten and selenium atoms, respectively. The remaining parameters represent orbital onsite energies and crystal-field splittings. The hopping matrices $t^{\text{W-W}}$, $t^{\text{Se-Se}}$, and $t^{\text{W-Se}}$ are taken from Refs.~\cite{khoeini,silva-guill}.
\begin{figure}[h]
	\centering
	\includegraphics[width=0.48\textwidth]{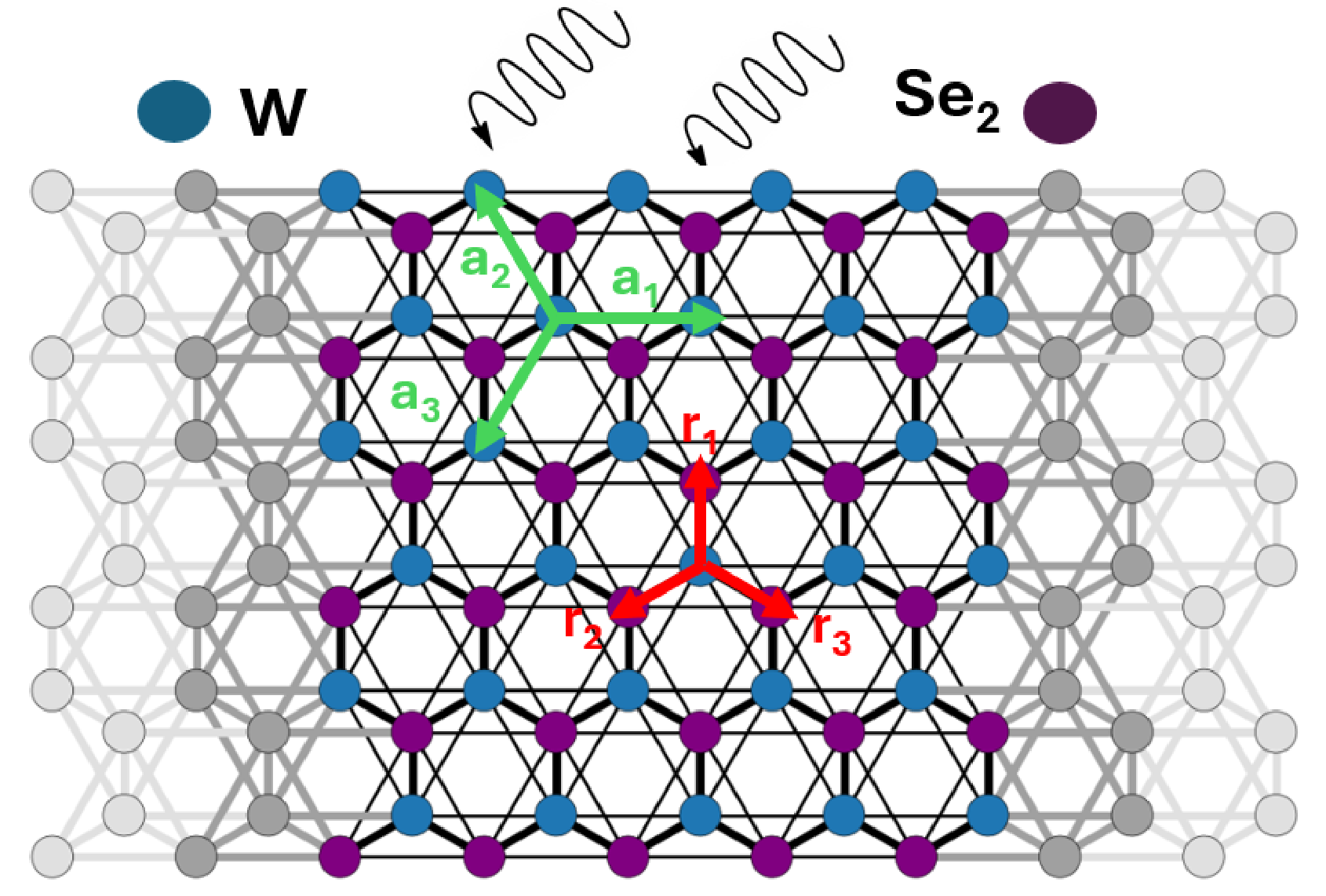}
	%\captionsetup{singlelinecheck=false, justification=justified}
	\caption{Schematic lattice structure of a monolayer WSe$_2$ device connected to two semi-infinite electrodes (gray), maintained at temperatures $T+\frac{\Delta T}{2}$ and $T-\frac{\Delta T}{2}$, where $\Delta T$ denotes an infinitesimal temperature bias. Blue (purple) circles represent W (Se) atoms. The vectors ${\bf a}_i$ (green arrows) denote nearest-neighbor intra-sublattice hopping (W-W and Se-Se), while ${\bf r}_i$ (red arrows) indicate nearest-neighbor inter-sublattice hopping (W-Se). The black wave illustrates incident light irradiation perpendicular to the monolayer surface.}
	\label{fig:sys}
\end{figure}
Figure~\ref{fig:sys} illustrates the schematic of the system under consideration. It consists of a zigzag-edged monolayer WSe$_2$ nanoribbon of finite width connected to two semi-infinite leads. The nanoribbon lattice is composed of tungsten (W) and selenium (Se) atoms, represented by distinct colored sites, arranged in a hexagonal crystal structure characteristic of transition metal dichalcogenides. 
%The ribbon is terminated by inequivalent atoms along the transverse ($y$) direction, giving rise to zigzag edges, while translational invariance is preserved along the longitudinal ($x$) direction.
%%%%%%%%%%%%

Table~\ref{tab:TBparams} summarizes the SOC strengths and tight-binding parameters used in this work. The quantities $\lambda_W$ and $\lambda_{Se}$ denote the SOC constants of W and Se atoms, respectively. The on-site crystal-field parameters $\Delta_0$ and $\Delta_2$ represent the energy levels of the symmetry-distinct orbitals retained in the model. In the trigonal prismatic crystal field, the metal $d$-orbitals split into a singlet $\left(d_{3z^2-r^2}\right)$ and a doublet $\left(d_{x^2-y^2}, d_{xy}\right)$, whose on-site energies are described by $\Delta_0$ and $\Delta_2$, respectively. The parameters $\Delta_p$ and $\Delta_z$ denote the on-site energies of the chalcogen in-plane $\left(p_x,p_y\right)$ and out-of-plane $\left(p_z\right)$ orbitals. Electron hopping between different atomic sites is described using
Slater-Koster parameters: $V_{pd\sigma}$ and $V_{pd\pi}$ for metal-chalcogen
(W-Se) bonds, $V_{dd\sigma}$, $V_{dd\pi}$, and $V_{dd\delta}$ for metal-metal
(W-W) interactions, and $V_{pp\sigma}$ and $V_{pp\pi}$ for chalcogen-chalcogen
(Se-Se) couplings.

\begin{table}[t]
	\centering
	\caption{Slater-Koster parameters \cite{silva-guill} and SOC parameters
		\cite{kosmider} used in the tight-binding calculations (in eV) for
		monolayer WSe$_2$.}
	\label{tab:TBparams}
	\begin{tabular*}{\columnwidth}{@{\extracolsep{\fill}} l c r}
		\hline\hline
		& Parameter & WSe$_2$ \\
		\hline
		W-Se
		& $V_{pd\sigma}$ & $5.803$\\
		& $V_{pd\pi}$ & $-1.081$ \\
		\hline
		W-W
		& $V_{dd\sigma}$ & $-1.129$ \\
		& $V_{dd\pi}$ & $0.094$ \\
		& $V_{dd\delta}$ & $0.317$ \\
		\hline
		Se-Se
		& $V_{pp\sigma}$ & $1.530$ \\
		& $V_{pp\pi}$ & $-0.123$ \\
		\hline
		SOC
		& $\lambda_{W}$ & $0.251$ \\
		& $\lambda_{Se}$ & $0.439$ \\
		\hline
		Crystal fields
		& $\Delta_{0}$ & $-0.935$ \\
		& $\Delta_{2}$ & $-2.321$ \\
		& $\Delta_{p}$ & $-5.629$ \\
		& $\Delta_{z}$ & $-6.759$ \\
		\hline\hline
	\end{tabular*}
\end{table}
%%%%%%%%%%%%%%%%%%%%%%%%%%%%%%%%%%%%%%%%%%%%%%

To investigate quantum transport and thermoelectric properties, the nanoribbon is symmetrically coupled to two semi-infinite leads at its left and right ends. The leads share the same lattice structure and tight-binding parameters as the central scattering region, ensuring minimal contact resistance and eliminating spurious interface effects. The left and right electrodes are maintained at slightly different temperatures, $T + \Delta T/2$ and $T - \Delta T/2$, respectively, where $\Delta T$ is an infinitesimally small temperature bias. This configuration allows us to restrict our analysis to the linear-response regime. Electronic transport through the system is treated within the coherent transport framework by solving the scattering problem using the wave-function matching approach as implemented in the \textsc{Kwant} package\cite{groth2014kwant}.

\subsection{Incorporation of Light Irradiation}
Light irradiation is incorporated through a time-dependent vector potential within the minimal-coupling and Floquet-Bloch formalism~\cite{PhysRevLett.110.200403,sambe,grifoni1998driven,delplace}. The coupling is introduced in Eq.~(\ref{eq:1}) via
\begin{equation}
\mathbf{k} \rightarrow \mathbf{k} + \frac{q}{\hbar}\mathbf{A}(\tau),
\end{equation}
where $q$ is the electronic charge. Within the dipole approximation, valid when the radiation wavelength exceeds the system size, the vector potential is spatially uniform and can be written as
\begin{equation}
\mathbf{A}(\tau)=
\left(
\mathcal{A}_x \sin\Omega\tau,\;
\mathcal{A}_y \sin(\Omega\tau+\theta),\;
0
\right),
\end{equation}
allowing linear, circular, or elliptical polarization depending on $\mathcal{A}_x$, $\mathcal{A}_y$, and $\theta$.

We define the dimensionless driving strengths
\begin{equation}
A_x=\frac{e\mathcal{A}_x a}{\hbar},
\qquad
A_y=\frac{e\mathcal{A}_y a}{\hbar},
\end{equation}
with lattice constant $a$. For $A_{x,y}\sim \mathcal{O}(1)$, the corresponding vector potential is $\sim10^{-6}\,\text{T}\cdot\text{m}$. Since the wavelength considered ($\sim300\,\text{nm}$) is much larger than the device size (few tenths of nm), the dipole approximation is well justified.

Under periodic driving, the nearest-neighbor hopping between sites $n$ and $m$ acquires Floquet components~\cite{PhysRevLett.110.200403,kallol,GANGULY2021302}:
\begin{equation}
\tilde{t}_{nm}^{p,q}
=
\frac{t_{nm}}{\mathbb{T}}
\int_0^{\mathbb{T}}
e^{i\Omega\tau(p-q)}
e^{i\mathbf{A}(\tau)\cdot\mathbf{d}_{nm}}
\, d\tau,
\label{effhop}
\end{equation}
where $\mathbf{d}_{nm}=d_x\hat{\mathbf{x}}+d_y\hat{\mathbf{y}}$ and $\mathbb{T}=2\pi/\Omega$.

Evaluating the time average yields
\begin{equation}
\tilde{t}_{nm}^{p,q}
=
t_{nm}
e^{i(p-q)\Theta}
J_{(p-q)}(\Gamma),
\label{effhop1}
\end{equation}
with
\begin{align}
\Gamma &= \sqrt{(A_x d_x)^2+(A_y d_y)^2
+2A_xA_y d_x d_y\cos\theta}, \\
\Theta &= \tan^{-1}
\left(
\frac{A_y d_y \sin\theta}
{A_x d_x + A_y d_y \cos\theta}
\right).
\end{align}
Here $J_{(p-q)}$ is the Bessel function of order $(p-q)$. The resulting hopping becomes bond-direction dependent, leading to light-induced anisotropy in the lattice~\cite{delplace,harish}.

\subsection{Electronic Thermoelectric Quantities} 
The two-terminal transmission probability ${\mathcal T}(E)$ is calculated numerically using the \texttt{Kwant} package. The thermoelectric properties of the system are then evaluated within the Landauer-B\"uttiker formalism. In the linear-response regime, the electrical conductance $G$, Seebeck coefficient $S$, and electronic thermal conductance $\kappa_{\mathrm{el}}$ are obtained from the Landauer integrals~\cite{finch2009,zerah,kallol}.
\begin{subequations}
	\label{eq:transport}
	\begin{align}
		G &= \frac{2e^2}{h} L_0,
		\label{eq:transport:a} \\[4pt]
		S &= -\frac{1}{eT}\frac{L_1}{L_0},
		\label{eq:transport:b} \\[4pt]
		\kappa_{\mathrm{el}} &=
		\frac{2}{hT}
		\left(L_2 - \frac{L_1^2}{L_0}\right).
		\label{eq:transport:c}
	\end{align}
\end{subequations}

where the Landauer integrals are defined as
\begin{equation}
	L_n = -\int_{-\infty}^{\infty}
	\mathcal{T}(E)\,(E-E_F)^n\,
	\frac{\partial f(E)}{\partial E}\, \text{d}E.
	\label{eq:Ln}
\end{equation}
Here $\mathcal{T}$, $E_F$, and $f(E)$ represent the two-terminal transmission probability, Fermi energy, and Fermi-Dirac distribution function, respectively.

The thermoelectric efficiency is quantified by the dimensionless figure of merit
\begin{equation}
	ZT = \frac{GS^2T}{\kappa},
	\label{eq:ZT}
\end{equation}
where $\kappa$=$\kappa_{\text{el}}+\kappa_{\text{ph}}$ includes both electronic and phononic contributions to thermal conductance. While all electronic transport coefficients are obtained from the Landauer formalism, the phonon thermal conductance $\kappa_{ph}$ is evaluated separately and discussed in the following section.

First-principles calculations were carried out within the framework of density functional theory (DFT), employing the Quantum Espresso (QE) package~\cite{dft}. Perdew-Burke-Ernzerhof (PBE)~\cite{pbe1} functional using the generalized gradient approximation (GGA)~\cite{pbe2} technique was employed to describe the exchange-correlation interactions in all calculations performed using the QE package. Fully relativistic pseudopotentials with the PBE functional were employed to account for spin-orbit coupling effects. For structural optimization, a kinetic energy cutoff of 60 Ry was used for the wave functions, while a cutoff of 240 Ry was applied to the charge density and the same cutoff values were consistently applied to both electronic and phonon calculations. A vacuum gap of 15$\,$\AA~is introduced along the $z$-axis to eliminate van der Waals (vdW) interactions between periodic layers. The lattice parameters were fully optimized, and atomic positions were relaxed until the total energy convergence threshold reached $10^{-10}$ Ry. The optimized crystal structure was used to perform self-consistent calculations with an 8$\times$8$\times$1 Monkhorst-Pack $k$-mesh~\cite{wang2021}. For the evaluation of the density of states (DOS), non-self-consistent field (NSCF) calculations were performed with a denser Monkhorst-Pack~\cite{wang2021} grid of 24$\times$24$\times$1, and the tetrahedron method was adopted as the smearing scheme to ensure accurate Brillouin zone sampling. We have obtained phonon dispersion using density functional perturbation theory (DFPT)~\cite{dfpt} along the $q$-point as 4$\times$4$\times$1 mesh in the Brillouin zone. The lattice thermal conductivity of superlattice was evaluated using the ShengBTE package~\cite{bte} which employs an iterative solution of the phonon Boltzmann transport equation. For the third-order IFCs, a finite displacement method was applied, in which three atoms within the supercell were displaced simultaneously, and the induced forces on the remaining atoms were calculated. A 4$\times$4$\times$1 supercell configuration was used to account for anharmonic interactions involving up to third-nearest neighbours, ensuring the accuracy of the phonon scattering processes. Within ShengBTE, the relevant intrinsic phonon scattering channels, namely Normal and Umklapp three-phonon processes, are explicitly considered.

\section{Results and discussion}

The monolayer WSe$_2$ adopts a honeycomb lattice with hexagonal symmetry, characterized by the space group $P\bar{6}m2 \left(D_{3h}\right)$, which is typical of group-VI transition metal dichalcogenide monolayers~\cite{ozbal}. In this structure, each Tungsten (W) atom is covalently coordinated to two Selenium (Se) atoms arranged in a trigonal prismatic geometry, with one Se atoms located in the upper sublayer and one in the lower sublayer. This coordination results in a single Se-W-Se sandwich layer, where the W atom occupies the central plane between the two Se planes, giving rise to strong intrelayer covalent bonding and high structural stability.

\begin{figure}[h]
	\centering
	\includegraphics[width=0.48\textwidth]{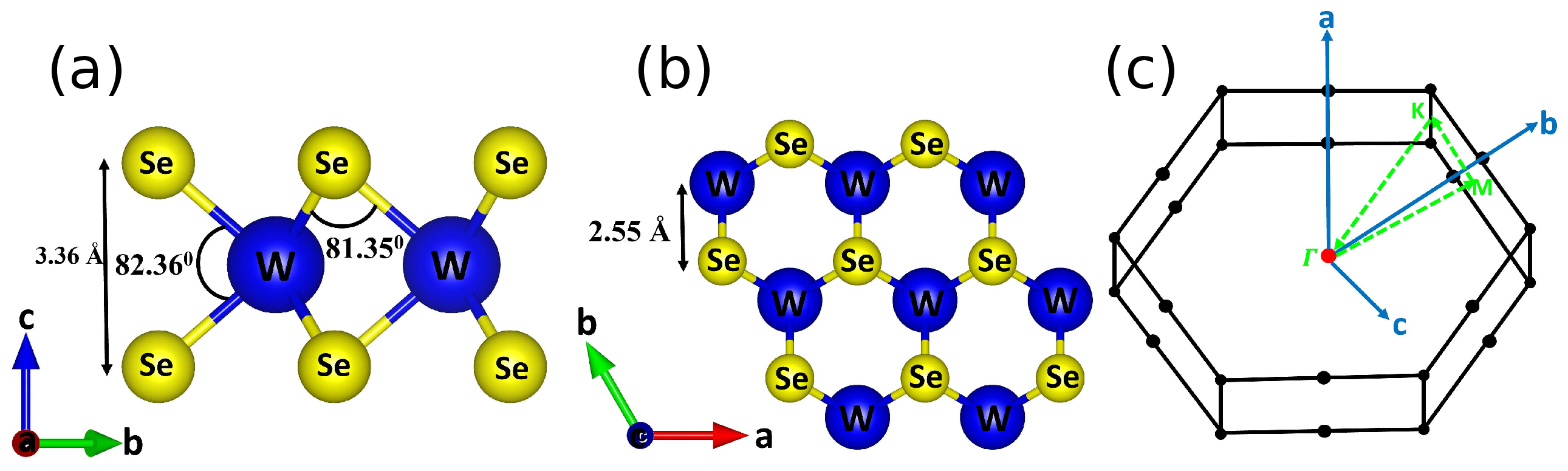}\vskip 0.1 in
		\includegraphics[width=0.48\textwidth]{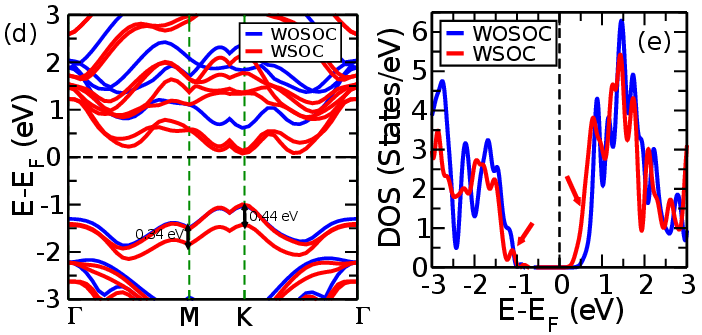}
	\caption{(a) Side view and (b) top view of the optimized crystal structure of monolayer WSe$_2$, illustrating the trigonal prismatic coordination of W and Se atoms. (c) The corresponding hexagonal Brillouin zone with high-symmetry points and the standard $\Gamma$-M-K-$\Gamma$ path used for electronic structure calculations. (d) Electronic band structures of monolayer WSe$_2$ calculated without spin-orbit coupling (WOSOC, blue lines) and with spin-orbit coupling (WSOC, red lines). Here the SOC-induced band splitting energies at high-symmetry points are highlighted. (e) Total electronic density of states (DOS) computed with and without SOC, where the red arrows highlight prominent DOS peaks induced by SOC near the band edges.}
	\label{raju1}
\end{figure}

The trigonal prismatic coordination distinguishes the 1H phase from the octahedral (1T) polymorph and plays a crucial role in determining the electronic and vibrational properties of the monolayer. Although bulk WSe$_2$ exhibits an ABABA$\ldots$ hcp-type stacking sequence due to interlayer interactions, the monolayer retains only a single Se-W-Se unit, eliminating interlayer coupling and thereby reducing the symmetry from $D_{6h}$ in the bulk to $D_{3h}$ in the two-dimensional limit. This symmetry reduction is responsible for the absence of inversion symmetry in the monolayer, which has important consequences for valley dependent phenomena. The optimized structural parameters, including W-Se bond lengths and Se-W-Se bond angles, are consistent with previously reported first-principles calculations, confirming the reliability of the relaxed geometry as shown in Figs.~\ref{raju1}(a) and (b). The in-plane lattice constant is found to be $a=3.32\,$\AA, in good agreement with previously reported experimental values~\cite{easy,boccuni}, while the out-of-plane lattice parameter, defined by the vacuum-separated supercell, is $c=15.06\,$\AA, ensuring negligible interaction between periodic images. The effective monolayer thickness, measured as the vertical distance between the uppermost and lowermost Se atoms, is $h=3.36\,$\AA. These values are in close agreement with available experimental and theoretical reports, validating the structural model employed in this study~\cite{easy,zulfi,affandi}. The corresponding hexagonal Brillouin zone shown in Fig.~\ref{raju1}(c) features the high-symmetry points $\Gamma$, M, and K, which define the standard $\Gamma$-M-K-$\Gamma$ path employed for electronic structure calculations.

The electronic band structures of monolayer WSe$_2$, calculated with and without spin-orbit coupling (SOC) using the optimized lattice parameters, are presented in Fig.~\ref{raju1}(d) and show good agreement with previously reported theoretical studies~\cite{tang,amin,kxchen}. In the absence of SOC, WSe$_2$ exhibits a direct band gap of 1.63$\,$eV, with both the conduction band minimum (CBM) and valence band maximum (VBM) located at the K point of the Brillouin zone. Upon inclusion of SOC, a pronounced energy-level splitting emerges due to the coupling between the electron's spin and orbital angular momentum, resulting in a reduced band gap of 1.33$\,$eV maintaining band curvature unaltered. The spin-orbit interaction originates from the relativistic coupling between the spin and orbital magnetic moments of electrons and is described by the Hamiltonian $H_{\text{SOC}} =\alpha {\mathbf L} \cdot {\mathbf S}$, where $\alpha$ denotes the SOC constant and ${\mathbf L}$ and ${\mathbf S}$ represent the orbital and spin angular momentum operators, respectively. The strong spin-orbit interaction in monolayer WSe$_2$ originates from the strong relativistic effects associated with the intrinsically heavy Tungsten atom. SOC lifts the spin degeneracy of the electronic states and induces pronounced band splitting, particularly near the high-symmetry points of the Brillouin zone. At the $\Gamma$ point, SOC splits the degenerate states into a fourfold-degenerate $j=3/2$ band and a twofold-degenerate $j=1/2$ band, with the corresponding energy levels given by $E_- = E_l - 2\lambda_l$ and $E_+ = E_l + 2\lambda_l$, where $\lambda_l$ denotes the SOC strength~\cite{blume}. The resulting SOC-induced energy separation yields effective coupling strengths of approximately 0.34$\,$eV at the M point and 0.44$\,$eV at the K point. These results underscore the significant role of spin-orbit interaction in shaping the electronic structure of monolayer WSe$_2$ and consistent with prior theoretical reports~\cite{latzke}.

To further elucidate the influence of spin-orbit coupling on the electronic structure, the total density of states with and without spin-orbit coupling is calculated and shown in Fig.~\ref{raju1}(e). The inclusion of SOC leads to a noticeable redistribution of electronic states near the band edges. In particular, the SOC induced band splitting results in an enhanced DOS in the conduction band region, giving rise to a pronounced peak located around 0.5$\,$eV above the Fermi level. This enhancement originates from the lifting of band degeneracy and the increased density of available electronic states within a narrow energy range. Such modifications in the DOS have direct implications for thermoelectric transport. Within the Boltzmann transport formalism, the Seebeck coefficient $S$ is strongly influenced by the energy dependence of the density of states near the Fermi level, with larger and more rapidly varying DOS generally leading to enhanced thermopower. Consequently, the elevated conduction band DOS induced by SOC is expected to favor an increase in the Seebeck coefficient, highlighting the important role of spin-orbit interaction in tuning the thermoelectric response of monolayer WSe$_2$.

%%%%%%%%%%%%%%%%%%%%%%%%%%%%%%%%%%%%%%%%% Transmission %%%%%%%%%%%%%%%%%%%%%%%%%%%%%%%%%%%%%%%%%%%
Because the driving field is time-periodic, the system can be treated within Floquet theory, where a $\mathbb{D}$-dimensional driven lattice is mapped onto an effective static problem in $(\mathbb{D}+1)$ dimensions~\cite{PhysRevLett.110.200403,sambe,grifoni1998driven}. The Bloch spectrum is thereby replicated into Floquet sidebands separated by $\hbar\Omega$, corresponding to photon-dressed copies of the original system.

The degree of hybridization between Floquet sectors is governed by the ratio of the driving frequency to the intrinsic electronic bandwidth $BW$. In the high-frequency regime ($\hbar\Omega \gg BW$), inter-sector coupling is weak and the dynamics are dominated by the zeroth Floquet sector ($p=q=0$) in Eq.~(\ref{effhop1}), while higher-order photon processes are suppressed. In contrast, for $\hbar\Omega \lesssim W$, strong mixing between Floquet replicas significantly modifies the quasienergy spectrum and may degrade coherent transport. 

For monolayer WSe$_2$, the relevant multi-orbital bandwidth is approximately $BW\sim 20\,\text{eV}$. We therefore restrict our analysis to the high-frequency regime $\hbar\Omega > BW \approx 20\,\text{eV}$. In the present work, we choose $\hbar\Omega=20\,\text{eV}$, corresponding to $\Omega\sim3\times10^{16}\,\text{Hz}$ (UV regime), which satisfies the high-frequency condition.

For dimensionless driving strength of order unity, the corresponding field amplitudes are estimated as $E_0\sim 6.1\,\text{V/\AA}$, $B_0\sim 200\,\text{T}$, and intensity $I\sim 10^{10}\,\text{W/cm}^2$. These values represent effective model parameters capturing the relevant energy scales. In engineered systems with enlarged lattice spacing, the required fields are substantially reduced while preserving the same dimensionless coupling~\cite{delplace,artificial}. Since WSe$_2$ is non-magnetic, the dominant modification arises from the electric-field-induced renormalization of hopping amplitudes.

Having established the validity of the high-frequency approximation, we now proceed to discuss the resulting electronic and transport properties.

\begin{figure}[h]
	\centering
	\includegraphics[width=0.4\textwidth]{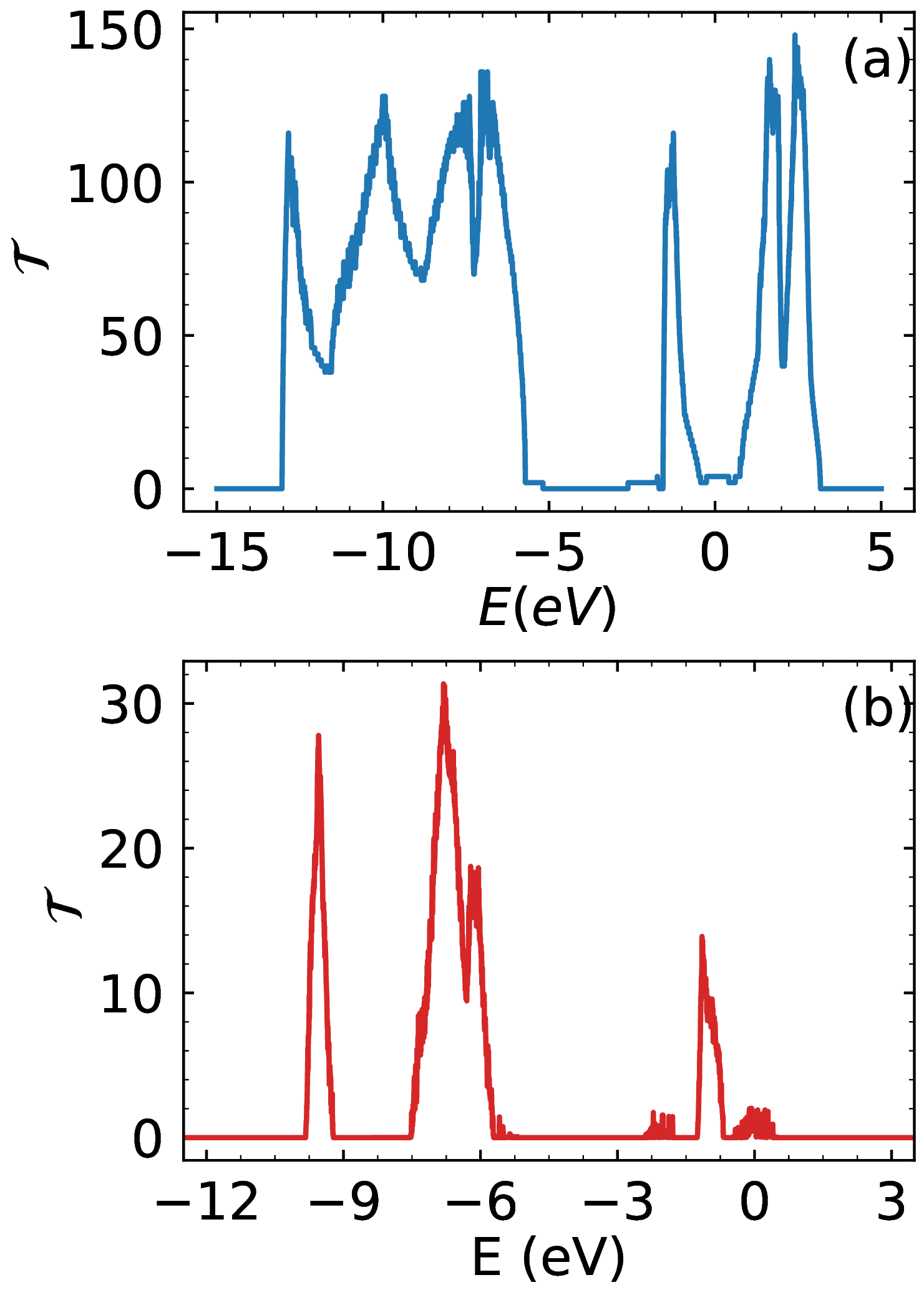}
	%\captionsetup{singlelinecheck=false, justification=justified}
	\caption{Transmission probability $\mathcal{T}(E) $ of a zigzag monolayer WSe$_2$ nanoribbon as a function of energy in the (a) absence and (b) presence of light. The light parameters are  A$_x = 1$, A$_y= 2$, and $\theta=\pi/4$.}
	\label{fig:transm}
\end{figure}

Figure~\ref{fig:transm} depicts the behavior of transmission probability as a function of energy for zigzag WSe$_2$. The length of the WSe$_2$ is fixed at $30~$nm and the width at $15~$nm. The widths of the electrodes are considered as same as that of the central system. In the absence of light, the transmission probability extends over a broad energy range and displays large, nearly continuous plateaus, indicating the presence of multiple propagating channels.
When irradiation is applied, the transmission spectrum is dramatically altered. The broad transmission plateaus are replaced by a series of narrow, well-separated resonant peaks, and the overall magnitude of the transmission is significantly reduced. Moreover, the energy window supporting finite transmission becomes much narrower. These features indicate that irradiation suppresses most conducting channels and renders transport highly energy selective.
This behavior arises from the light-induced renormalization of the electronic hopping parameters, which introduces anisotropy and modifies the band dispersion. The resulting sharp and strongly energy-dependent transmission profile is particularly favorable for thermoelectric applications, as it enhances the energy filtering of charge carriers and can lead to an increased Seebeck coefficient, as discussed in the following sections.

%%%%%%%%%%%%%%%%%%%%%%%%%%%%%%%%%%%%%%%%% GSK %%%%%%%%%%%%%%%%%%%%%%%%%%%%%%%%%%%%%%%%%%%

\begin{figure*}[!htbp]
	\centering
	%\hspace*{-0.4cm}
	\includegraphics[width=1\textwidth]{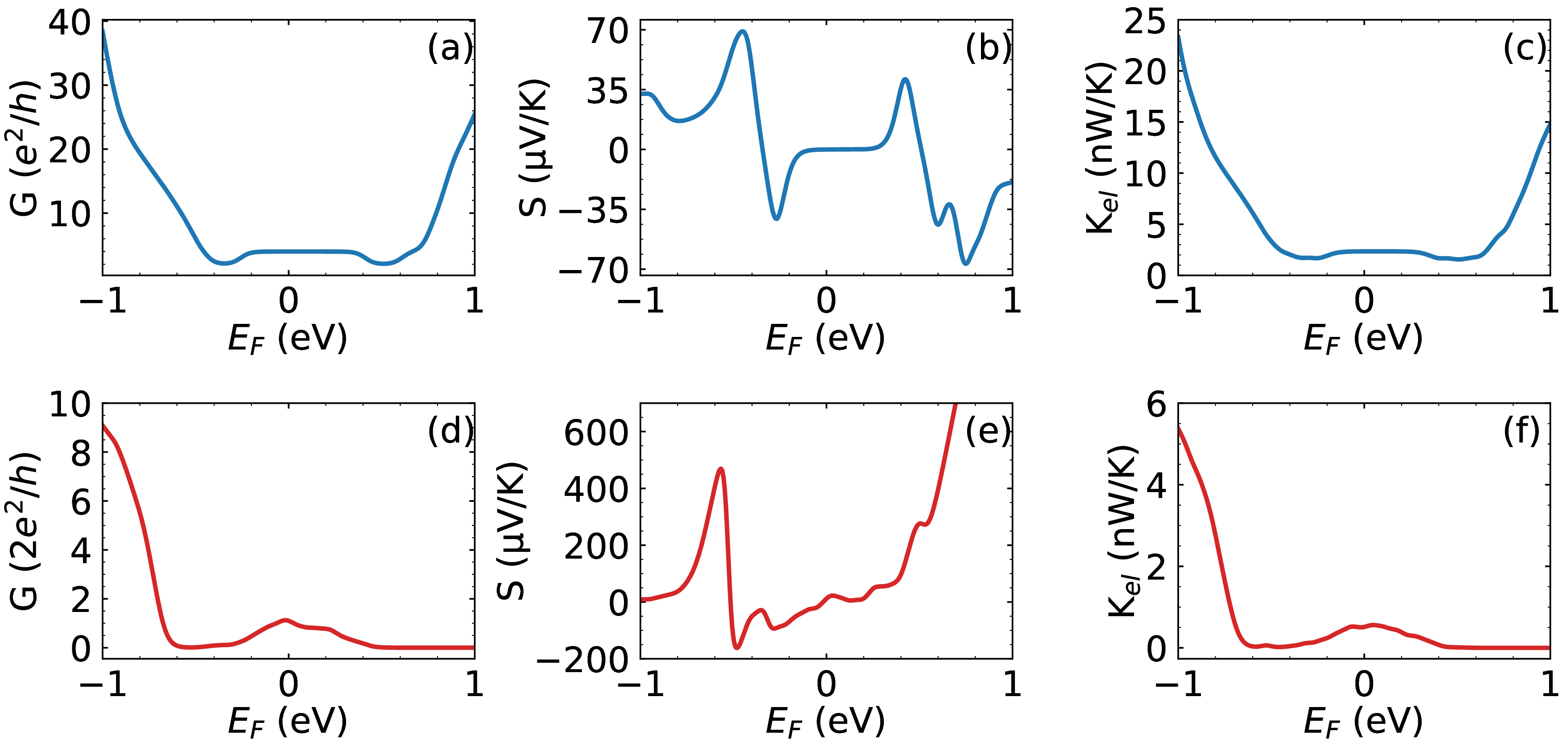}
	\vspace{-0.15cm}
	\caption{Behaviour of thermoelectric quantities as functions of the Fermi energy at room temperature ($T=300$ K). Panels (a)-(c) show the results without irradiation, while panels (d)-(f) correspond to the system under light irradiation. The electrical conductance $G$ (a, d), Seebeck coefficient $S$ (b,e), and electronic thermal conductance $\kappa_e$ (c, f) are calculated within the linear-response regime using Eq.~(\ref{eq:Ln}). All structural and light parameters are the same as those used in Fig.~\ref{fig:transm}.}
	\label{fig:GSK}
\end{figure*}

Figure~\ref{fig:GSK} presents the various thermoelectric properties as functions of $E_F$. The top row (a-c) corresponds to the case in the absence of irradiation, while the bottom row (d-f) shows the results under light exposure. 

The variation of the electrical conductance $G$ with the Fermi energy is shown in Figs.~\ref{fig:GSK}(a) and (d) in the absence and presence of irradiation, respectively, and is calculated using Eqs.~(\ref{eq:transport:a}) and (\ref{eq:Ln}). The conductance spectrum closely follows the $T-E$ curve shown in Fig.~\ref{fig:transm}. $G$ is plotted within a narrow energy range from $-1$ to $1$. In the absence of light, the conductance remains relatively large over a broad energy range and varies smoothly with $E_F$. This behavior reflects the presence of multiple open transmission channels and the broadband transport characteristics observed in Fig.~\ref{fig:transm}(a).
Upon introducing irradiation, the magnitude of $G$ is strongly suppressed across the entire energy window by a factor of 4, compared to the case without light. 

\begin{figure*}[!htbp]
	\centering	\includegraphics[width=0.33\textwidth,height=0.25\textwidth]{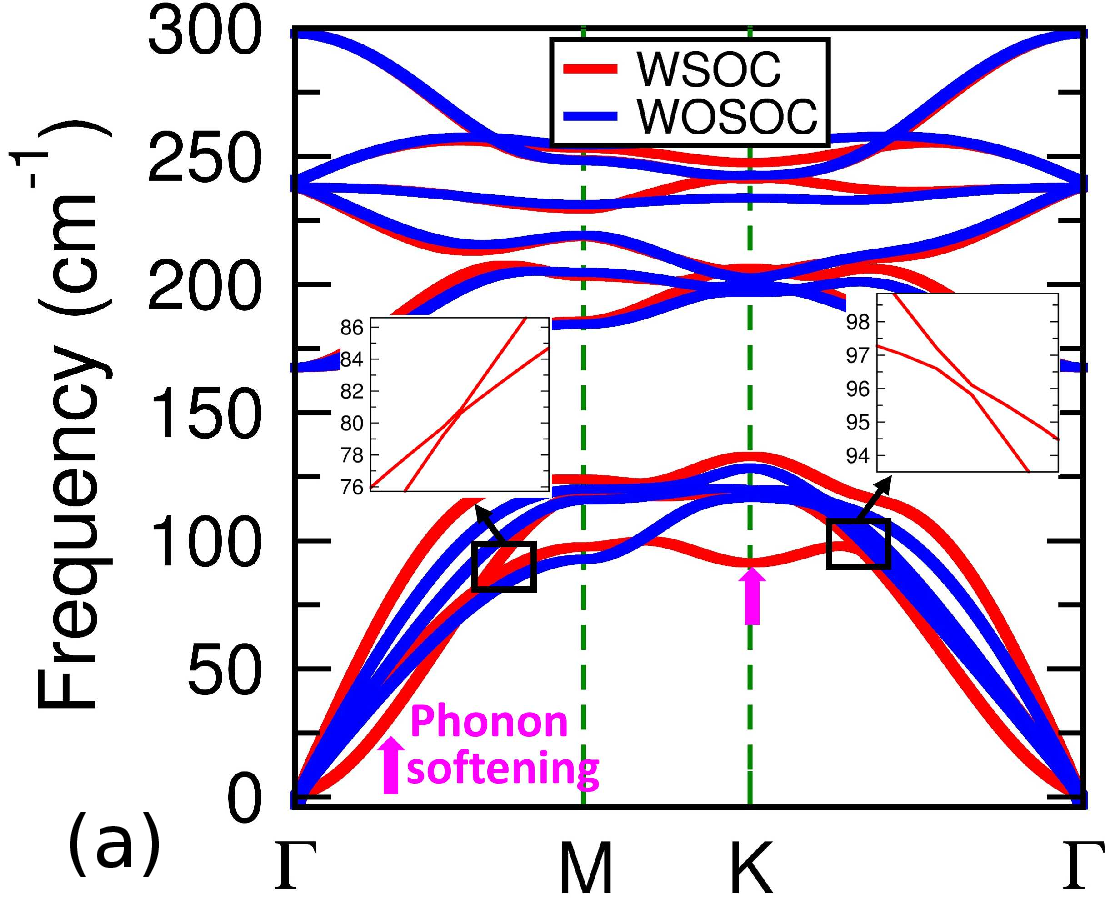}\hfill	\includegraphics[width=0.33\textwidth,height=0.25\textwidth]{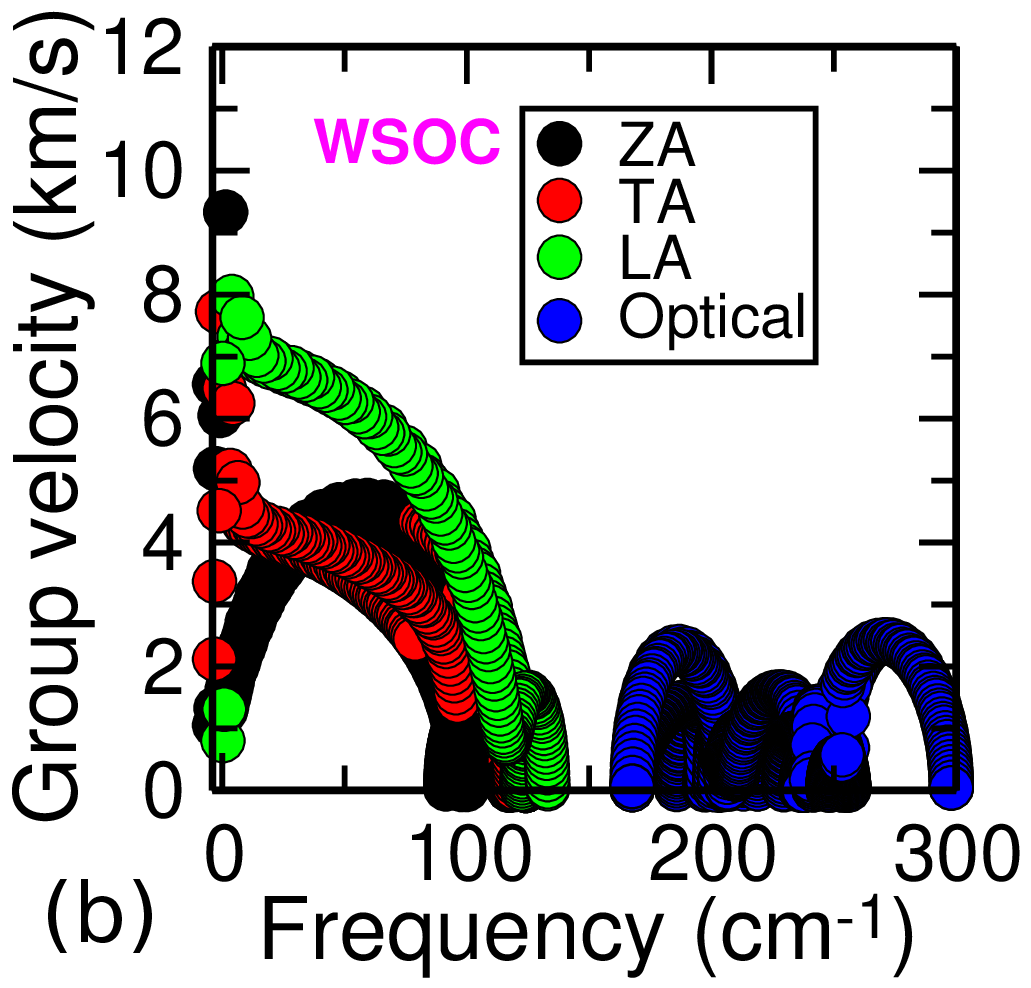}\hfill		\includegraphics[width=0.33\textwidth,height=0.25\textwidth]{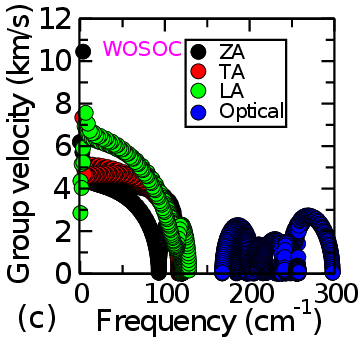}
	\caption{(a) Phonon dispersion of the system calculated without spin-orbit coupling (WOSOC, blue lines) and with spin-orbit coupling (WSOC, red lines) along the high-symmetry $\Gamma$-M-K-$\Gamma$ path. The black rectangular box highlights the SOC-induced avoided crossing between acoustic phonon branches, while the inset provides a magnified view of this region to clearly illustrate the avoided crossing. Phonon softening at the K and $\Gamma$ points is highlighted by magenta arrows. (b) Phonon group velocity as a function of frequency for the WSOC case, and (c) the corresponding group velocity for the WOSOC case, resolved into ZA, TA, LA, and optical modes.}
	\label{raju2}
\end{figure*}
The Seebeck coefficient $S$ is evaluated using Eq.~(\ref{eq:transport:b}), with its corresponding thermal integral, $L_1$, derived from  Eq.~(\ref{eq:Ln}).
In the absence of irradiation, the thermopower of the Wse$_2$ nanoribbon remains relatively small, as shown in Fig.~\ref{fig:GSK}(b), with typical values not exceeding $\sim 60\,\mu \text{V/K}$ over the considered Fermi energy range. Such moderate values are insufficient for efficient thermoelectric applications. This limited thermopower originates from the nearly symmetric transmission spectrum around the Fermi energy, as seen in Fig.~\ref{fig:transm}(a), which leads to partial cancellation of electron and hole contributions in the thermal integral $L_1$. Around $E_F = 0$, the thermopower is relatively small and fluctuates weakly about zero.
In sharp contrast, the thermopower under irradiation as shown in Fig.~ \ref{fig:GSK}(e) is dramatically enhanced  by more than an order of magnitude, reaching peak values of approximately $600~\mu\text{V/K}$. This substantial enhancement is particularly significant in view of the quadratic dependence of the thermoelectric figure of merit on the Seebeck coefficient, i.e. $ZT \propto S^2$. The origin of this behavior can be directly traced to the irradiation-induced modification of the electronic transmission spectrum. As illustrated in Fig.~\ref{fig:transm}(b), irradiation strongly renormalizes the hopping amplitudes, leading to sharp resonant features and pronounced asymmetry of the transmission function around the edges of the irradiation-induced gap. This asymmetry is crucial for generating a large thermopower, as the factor $(E - E_F)\partial f/\partial E$ appearing in $L_1$ selectively amplifies contributions from one side of the Fermi energy when $T(E)$ is no longer symmetric.
The electronic thermal conductance $\kappa_{\text{e}}$ exhibits a trend closely correlated with that of the electrical conductance. In the absence of light, $\kappa_{\text{e}}$, shown in Fig.~\ref{fig:GSK}(c), attains relatively large values, reaching up to $\sim 25\,$nW/K near the edges of the given Fermi energy window and remains finite around $E_F = 0$. This behavior reflects the broad energy window of finite transmission in the non-irradiated system. Under irradiation, however, $\kappa_{\text{e}}$ is significantly suppressed across the entire Fermi energy range, as shown in Fig.~\ref{fig:GSK}(f), with values dropping below $\sim 5~\text{nW/K}$. This suppression arises from the narrowing of the transmission window and the elimination of extended conducting channels, consistent with the transmission spectra in Fig.~\ref{fig:transm}(b).
The simultaneous enhancement of the thermopower and reduction of the electronic thermal conductance under irradiation constitutes a highly favorable scenario for thermoelectric performance. While the electrical conductance is reduced, the strong asymmetry in the transmission spectrum and the suppression of heat-carrying electronic states lead to a substantial increase in thermoelectric efficiency. These results demonstrate that electromagnetic irradiation provides an effective and reversible route for engineering the thermoelectric response of WSe$_2$ nanostructures through direct control of the energy-dependent transmission function.

%%%%%%%phonon%%%%%%%%%%%%%
%\section{Phonon dispersion}
To study the impact of spin-orbit coupling on lattice dynamics, we systematically investigated the phonon dispersion of monolayer WSe$_2$. Owing to the presence of three atoms per unit cell (one W and two Se), the phonon spectrum consists of nine vibrational modes, comprising three acoustic and six optical branches, in accordance with the 3N phonon degrees of freedom. As shown in Figure~\ref{raju2}(a), the overall phonon dispersions obtained from with SOC (WSOC) and without SOC (WOSOC) calculations exhibit nearly identical features, confirming that SOC does not significantly modify the dynamical stability. However, a discernible SOC-induced shift is observed in the low-frequency acoustic region, particularly for the out-of-plane flexural (ZA) mode. Upon inclusion of SOC, the ZA branch shows a slight softening near the $\Gamma$ point and at the K point (highlighted by magenta arrow in Fig.~\ref{raju2}(a)), while the longitudinal (LA) and transverse (TA) acoustic modes remain largely unaffected. This selective sensitivity of the ZA mode arises from its strong dependence on long-range interatomic force constants and out-of-plane restoring forces, which can be weakly altered by SOC-mediated changes in electronic structure~\cite{lindsay2010}. Similar SOC driven phonon softening effects have been reported in heavy-element systems such as PbTe and SnSe, where relativistic interactions modify bonding characteristics and reduce acoustic phonon frequencies, especially for lattice-flexible modes~\cite{wu2019,ztian2012,hung2023}. Comparable behaviour has also been observed in two-dimensional materials containing heavy atoms, where SOC subtly influences flexural phonons without affecting the overall phonon topology~\cite{,mobaraki2019}. Consequently, the SOC-induced softening of the ZA mode reflects a subtle modification of lattice dynamics, while the overall phonon characteristics of the system remain largely preserved.

\begin{figure}[h]
	\centering	\includegraphics[width=0.23\textwidth,height=0.21\textwidth]{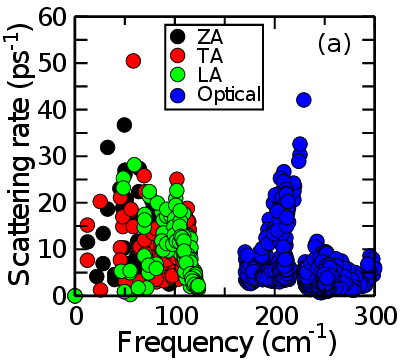}\hfill	\includegraphics[width=0.23\textwidth,height=0.21\textwidth]{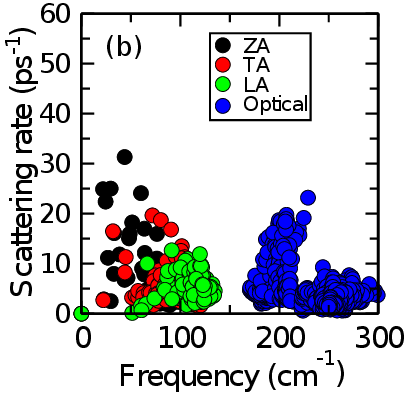}\vskip 0.1 in	\includegraphics[width=0.23\textwidth,height=0.21\textwidth]{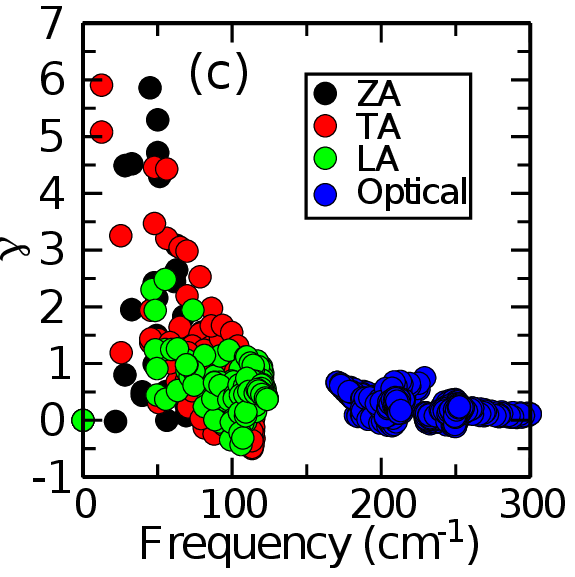}\hfill	\includegraphics[width=0.23\textwidth,height=0.21\textwidth]{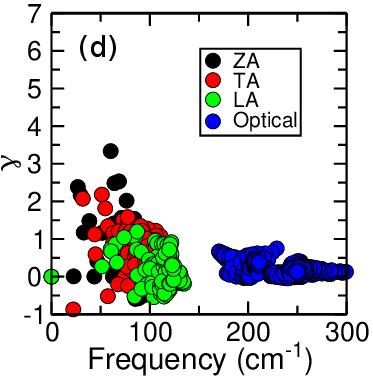}\vskip 0.1 in	\includegraphics[width=0.23\textwidth,height=0.21\textwidth]{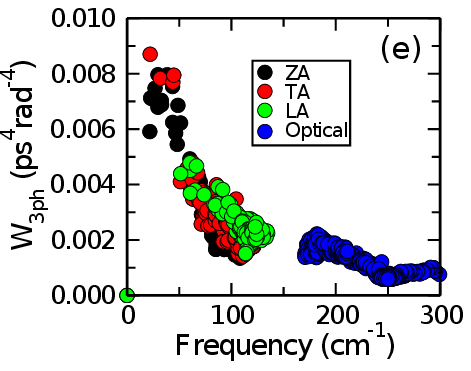}\hfill	\includegraphics[width=0.23\textwidth,height=0.21\textwidth]{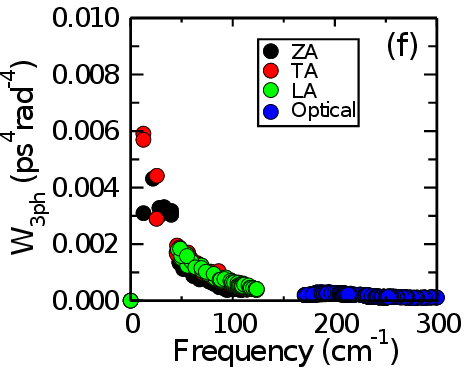}
	\caption{(a, b) Mode-resolved phonon scattering rates as a function of frequency for the with and without SOC systems, respectively. (c, d) Corresponding Grüneisen parameters ($\gamma$), highlighting the strength of phonon anharmonicity in both cases. (e, f) Three-phonon scattering phase space (W$_{3\text{ph}}$) for with and without SOC, respectively. Contributions from ZA, TA, LA, and optical phonon branches are shown using different colours.}
	\label{raju3}
\end{figure}
In addition to these frequency shift, SOC gives rise to avoided crossings between acoustic branches along the $\Gamma$-M and $\Gamma$-K directions. Such avoided-crossing behaviour enhances mode hybridization and increases acoustic phonon scattering by opening additional phonon-phonon interaction channels. Given that, heat transport in two-dimensional materials is predominantly governed by long-wavelength acoustic phonons, this SOC-induced enhancement of scattering is expected to contribute to the suppression of lattice thermal conductivity in monolayer WSe$_2$.

%\subsection{Thermal transport properties and lattice thermal conductivity}
{\bf Thermal transport properties and lattice thermal conductivity}
To gain deeper insight into phonon transport and to elucidate the mode-resolved origin of lattice thermal conductivity in monolayer WSe$_2$, we perform a comparative analysis of the phonon group velocities for calculations with spin-orbit coupling and without SOC. Phonon group velocity is defined as the derivative of the phonon frequency $\omega$ with respect to the wavevector $q$ and reflects the rate at which vibrational energy propagates through the lattice, given by~\cite{backman}, $v_{\lambda,q} = \dfrac{\partial \omega_{\lambda,q}}{\partial q}$, where, $\omega_{\lambda,q}$  denotes the phonon frequency. Figures~\ref{raju2}(b) and (c) illustrate, the group velocity distribution of phonon modes for with and without SOC, respectively. A clear reduction in group velocity is observed in the WSOC case, which directly correlates with the SOC-induced phonon softening evident in the dispersion curves (Fig.~\ref{raju2}(a)). In particular, the ZA phonon branch exhibits a pronounced softening upon inclusion of SOC, leading to substantially reduced group velocities compared to the WOSOC counterpart. This reduction strongly suppresses phonon energy transport and renders the WSOC system more susceptible to lower lattice thermal conductivity. Across both with and without SOC systems, optical phonon modes generally possess much smaller group velocities than acoustic modes and contribute marginally to heat transport. However, the additional suppression of group velocity of the ZA mode under SOC introduces an effective factor for phonon propagation. Consequently, the combined reduction in group velocities constitutes a key mechanism underlying the SOC-driven suppression of lattice thermal conductivity in monolayer WSe$_2$.

Although the phonon group velocities in both the cases are largely comparable, WSe$_2$ monolayer are largely comparable, the pronounced difference in lattice thermal transport originates primarily from variations in phonon scattering behaviour. As evident from the calculated scattering rates (Fig.~\ref{raju3}(a)), the WSOC system exhibits substantially enhanced phonon scattering compared to its WOSOC counterpart (Fig.~\ref{raju3}(b)). This increase is closely associated with SOC induced avoided crossings in the low-frequency region of the phonon dispersion, clearly visible in the inset of Fig.~\ref{raju2}(a). Such avoided crossings promote strong mode hybridization, particularly among the acoustic branches, thereby opening additional scattering channels. A more detailed mode resolved analysis (Fig.~\ref{raju3}(a)) reveals that the enhanced scattering in the WSOC system is dominated by the acoustic phonons, with the out-of-plane flexural (ZA) and transverse acoustic (TA) modes exhibiting markedly higher scattering rates below $\sim 100\,$cm$^{-1}$. In contrast, the optical phonon modes display nearly identical scattering rates in both with and without SOC cases, consistent with the negligible SOC induced changes observed in the optical branches of the phonon dispersion. These results indicate that SOC primarily amplifies acoustic phonon scattering through avoided crossing, which plays a decisive role in suppressing the lattice thermal conductivity of monolayer WSe$_2$.

The phonon scattering depends on two factors~\cite{xwu2016}: (i) the strength of each scattering channel depends on the anharmonicity of a phonon mode and is described by the Gr\"{u}neisen parameter ($\gamma$), which characterizes the relationship between phonon frequency and crystal volume change, $\gamma_i = -\dfrac{V}{\omega_i}\dfrac{\partial \omega_i}{\partial V}$. Here, $\omega_i$ and $V$ are the phonon frequency and crystal volume, respectively, and (ii) the number of channels available for a phonon to get scattered, which is determined by whether three phonon groups exist that can satisfy both energy and quasi-momentum conservations, and that can be characterized by the phase space for three-phonon processes~\cite{peng2016,lindsay}. As illustrated in Fig.~\ref{raju3}(c), the mode resolved Gr\"{u}neisen parameter ($\gamma$) of the WSOC system attains significantly larger values, reaching up to $\sim 6$, compared to WOSOC case (Fig.~\ref{raju3}(d)). This pronounced enhancement of $\gamma$ directly signals stronger lattice anharmonicity induced by spin-orbit coupling. Since the Gr\"{u}neisen parameter quantifies the sensitivity of phonon frequencies to lattice strain, higher $\gamma$ values imply an increased phonon-phonon interaction strength, which in turn intensifies anharmonic scattering and leads to a reduction in the lattice thermal conductivity ($\kappa_{\text{latt}}$). A closer inspection of the acoustic branches reveals that the enhancement of $\gamma$ in the WSOC system is predominantly associated with the out of plane flexural (ZA) and transverse acoustic (TA) modes, for which $\gamma$ exceeds $\sim 6$, whereas the corresponding values in the WOSOC system remain limited to approximately $\sim 4.5$. This mode selective increase reflects the greater susceptibility of low-frequency acoustic phonons to SOC induced modifications of the interatomic force constants and bonding characteristics. Given that, long-wavelength acoustic modes particularly ZA and TA plays a dominant role in heat transport of WSe$_2$, their elevated anharmonicity in the WSOC system provides a crucial microscopic mechanism for the enhanced phonon scattering and the consequent suppression of $\kappa_{\text{latt}}$.

Furthermore, besides group velocity, scattering rate and Gr\"{u}neisen parameter, the frequency dependent phase space of 3ph processes could also be a descriptor to emphasize the scattering mechanism clearly. Figures~\ref{raju3}(e) and (f) displaying the phase-space decomposition for 3ph scattering processes represent the absorption process ($\lambda + \lambda_1 \rightarrow \lambda_2$) and the emission process ($\lambda \rightarrow \lambda_1 + \lambda_2$) for with and without SOC, respectively.  The calculated phase space values show a corresponding trend, with SOC exhibiting the large phase space in the low frequency regime below 100$\,$cm$^{-1}$, implying a greater probability of phonon scattering and the presence of more available scattering channels. In comparison, the WOSOC system exhibits relatively low phase space contributions, predominantly below 100$\,$cm$^{-1}$, as illustrated in Fig.~\ref{raju3}(f). The combined effect of high anharmonicity and dense scattering phase space in the SOC induced system results in enhanced phonon scattering and the lowest $\kappa_{\text{latt}}$ among both systems.

\begin{figure}[h]
	\centering
	\includegraphics[width=0.35\textwidth]{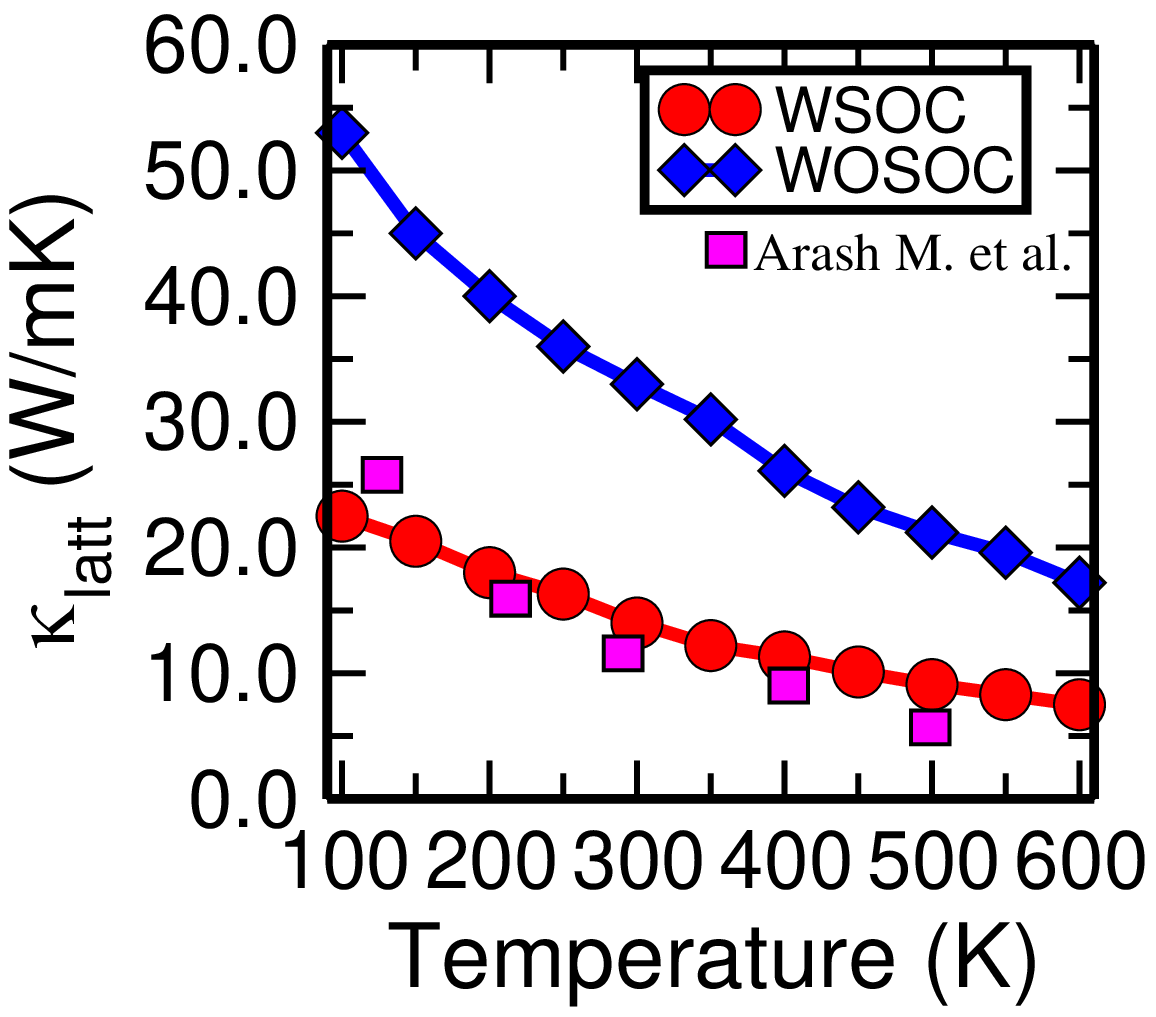}
	\caption{Calculated lattice thermal conductivity $\kappa_{\text{latt}}$ of monolayer WSe$_2$ as a function of temperature with and without spin-orbit coupling. Experimental data from prior reports are shown for comparison~\cite{mobaraki2019}, exhibiting close agreement with the SOC-included results.}
	\label{raju4}
\end{figure}
At last, we obtain the $\kappa_{\text{latt}}$ over the temperatures ranging from 100 to 600$\,$K for both the cases using the ShengBTE code and shown in Fig.~\ref{raju4}. The gradually decrease in $\kappa_{\text{latt}}$ with increasing temperature is attributed to heightened anharmonic phonon scattering rate and higher anharmonic scattering strength. The $\kappa_{\text{latt}}$ value calculated to be 15$\,$W/mK for the SOC-included case, whereas a value of 35$\,$W/mK is obtained without SOC. The predominance of more scattering channels, strong anharmonic scattering strength of acoustic modes and low group velocity are responsible in reducing $\kappa_{\text{latt}}$ in with SOC systems compare to WOSOC. Furthermore, a comparison with available experimental reports reveals that the WSOC results are in closer agreement with measured $\kappa_{\text{latt}}$ values~\cite{mobaraki2019}, thereby highlighting the crucial role of spin-orbit coupling in accurately capturing phonon transport properties and predicting the lattice thermal conductivity of monolayer WSe$_2$.

\begin{figure}[h]
	\centering
	\includegraphics[width=0.48\textwidth]
	{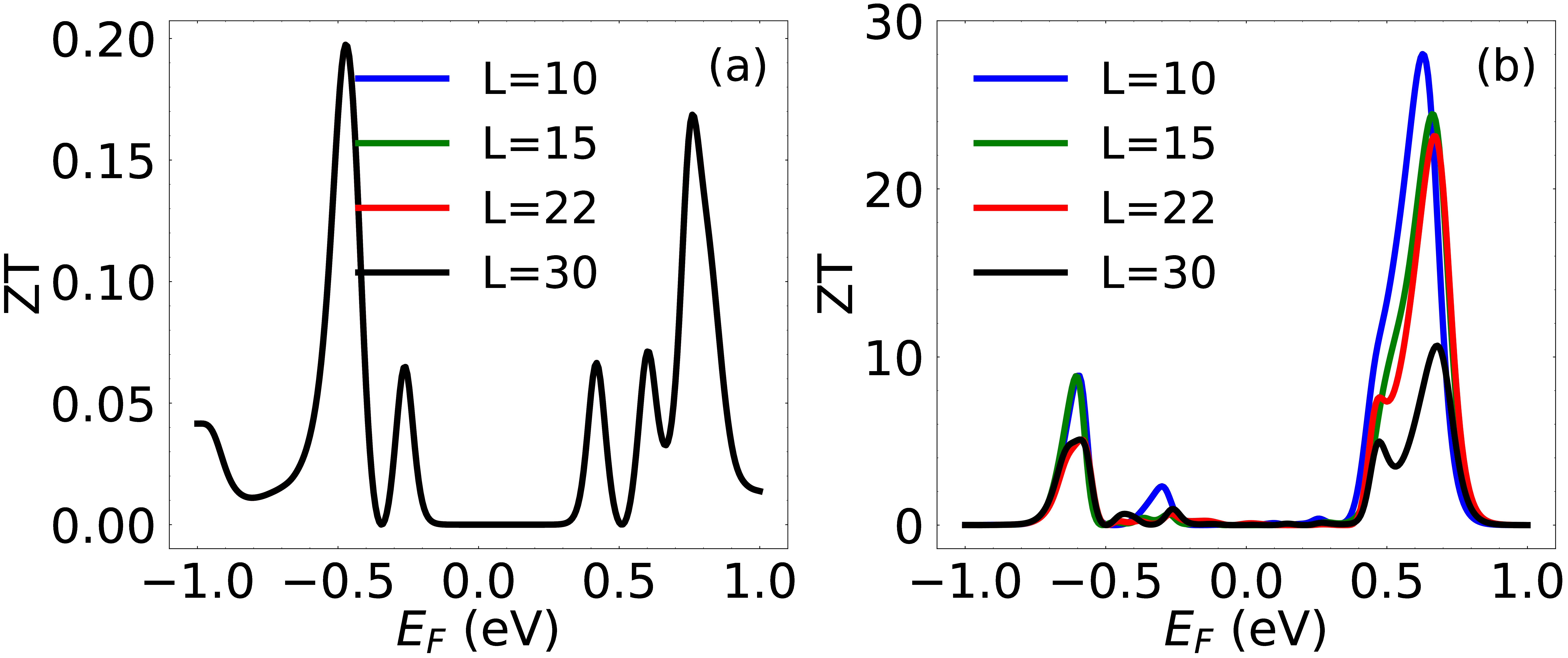}
	\caption{$ZT$ as a function of Fermi energy at room temperature ($T=300\,$K) for four different ribbon lengths: $10$, $15$, $22$, and $30\,$nm. Panels (a) and (b) correspond to the absence and presence of light, respectively. The ribbon width is fixed at $15\,$nm for all cases. All other system and light parameters are the same as those used in Fig.~\ref{fig:transm}.}
	\label{fig:ZT}
\end{figure}

%%%%%%%%%%%%%%%%%%%%%%%%%%%%%%%%%%%%%%%%% ZTmaxvsEF %%%%%%%%%%%%%%%%%%%%%%%%%%%%%%%%%%%%%%%%%%%
Having analyzed all the TE quantities derived from the tight-binding model and the phonon thermal conductivity from DFT analysis, we now evaluate the behavior of the FOM in the absence and presence of light at room temperature using Eq.~(\ref{eq:ZT}).
Figure~\ref{fig:ZT} shows $ZT$ as a function of $E_F$ for four different ribbon lengths, $L=10,15,22$, and 30$\,$nm while keeping the width fixed at $15\,$nm. The computed phonon thermal conductances for these lengths are 0.00225, 0.00337, 0.00495, and 0.00675$\,$pW/K, respectively, with the corresponding results represented by blue, green, red and black colors. Certainly, the phonon thermal conductance is significantly small compared to its electronic counterpart. All other physical and light parameters remain same with those mentioned in Fig.~\ref{fig:transm}.

In the absence of light, in Fig.~\ref{fig:ZT}(a), $ZT$ remains very small throughout the entire Fermi-energy range for all the considered ribbon lengths. This weak response comes from the large electronic transmission spectrum in Fig.~\ref{fig:transm}(a), which suppresses the Seebeck coefficient and limits $ZT$, despite finite electrical conductance.

Several key features emerge from Fig.~\ref{fig:ZT}(b) in the presence of light. $ZT$ exhibits highly favorable response.
The most prominent enhancement of $ZT$ occurs around $E_F \approx 0.6-0.8\,$eV. In this region, the maximum $ZT$ is about  28 for the smallest length and decreases slowly as we increase the length of the system. 
Importantly, the values of phonon thermal conductance $k_{\mathrm{ph}}$ remains extremely small for all ribbon lengths, which strongly suppress heat transport by lattice vibrations, thereby minimizing the denominator of the FOM thereby leading to the exceptionally large ZT values observed here.

We also observe a secondary peak around $E_F \approx -0.6\,$eV, where $ZT$ reaches values of order $ZT \sim 8$. Although smaller than the dominant high-energy peak, this demonstrates that both electron- and hole-like transport regimes can support enhanced thermoelectric performance under irradiation.
%%%%%%%%%%%%%%%%%%%%%%%%%%%%%%%%%%%%%%%%% ZTmaxvsW %%%%%%%%%%%%%%%%%%%%%%%%%%%%%%%%%%%%%%%%%%%
\begin{figure}[h]
	\centering
	\includegraphics[width=0.4\textwidth]{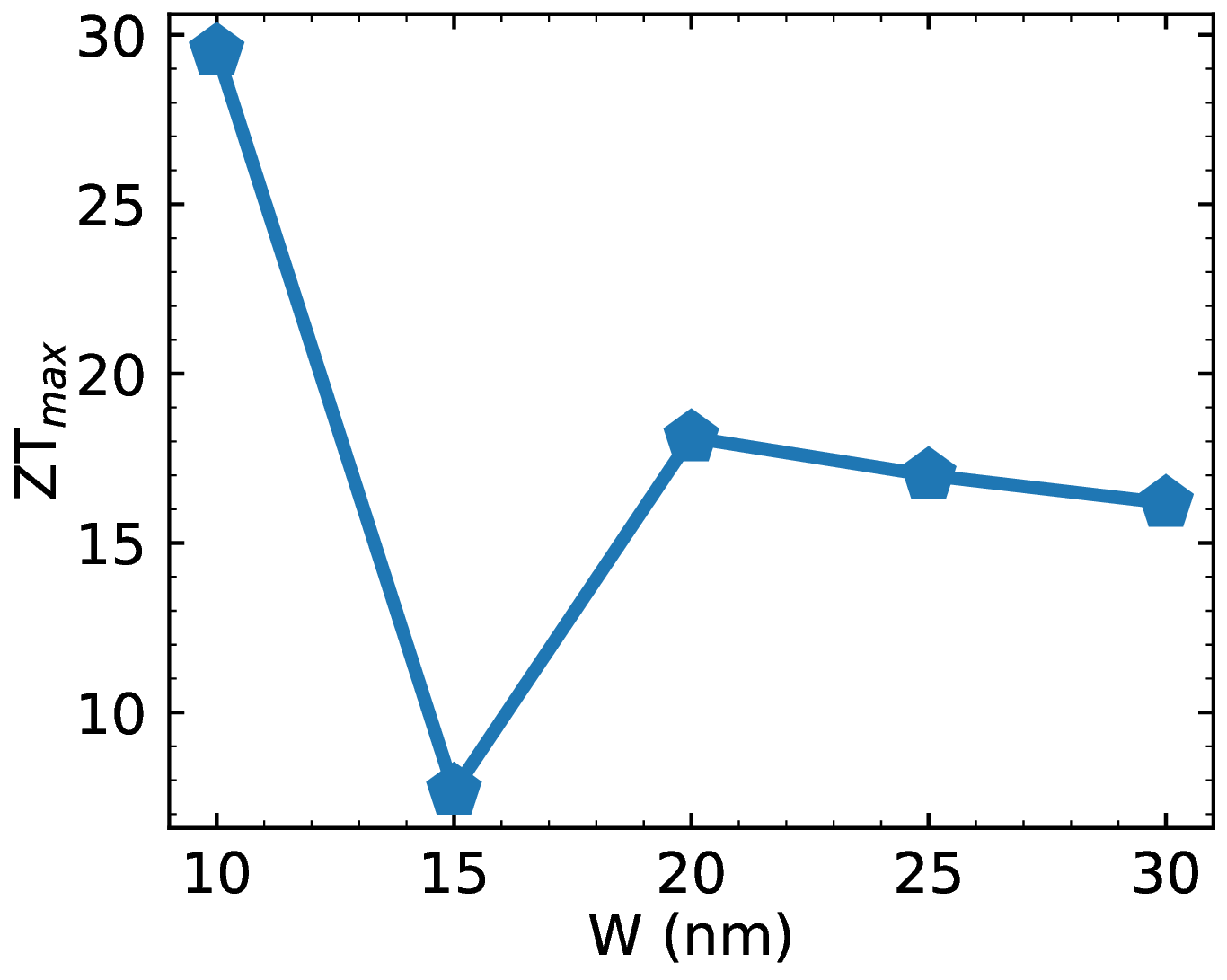}
	%\captionsetup{singlelinecheck=false, justification=justified}
	\caption{$ZT_{\mathrm{max}}$ as a function of width $W$ for irradiated WSe$_2$. The length is fixed at $30\,$nm for all cases. All other system parameters and the light parameters are the same as in Fig.~\ref{fig:transm}.}
	\label{fig:ZTvW}
\end{figure}

We further investigate how varying the width of the WSe$_2$ affects the behavior of $ZT$. To this end, we compute the maximum figure of merit, $ZT_{\mathrm{max}}$, obtained by optimizing $ZT$ over the Fermi energy window $-1$ to $1\,$eV as shown in Fig.~\ref{fig:ZTvW}. The results reveal a pronounced and nonmonotonic dependence of $ZT_{\mathrm{max}}$ on the ribbon width.
This behavior comes from the competition between geometric confinement and light-induced band renormalization, and indicates that high thermoelectric efficiency can be sustained over a broad range of nanoribbon widths.

%%%%%%%%%%%%%%%%%%%%%%%%%%%%%%%%%%%%%%%%% ZTmaxvsT %%%%%%%%%%%%%%%%%%%%%%%%%%%%%%%%%%%%%%%%%%%
\begin{figure}[h]
	\centering
	\includegraphics[width=0.42\textwidth]{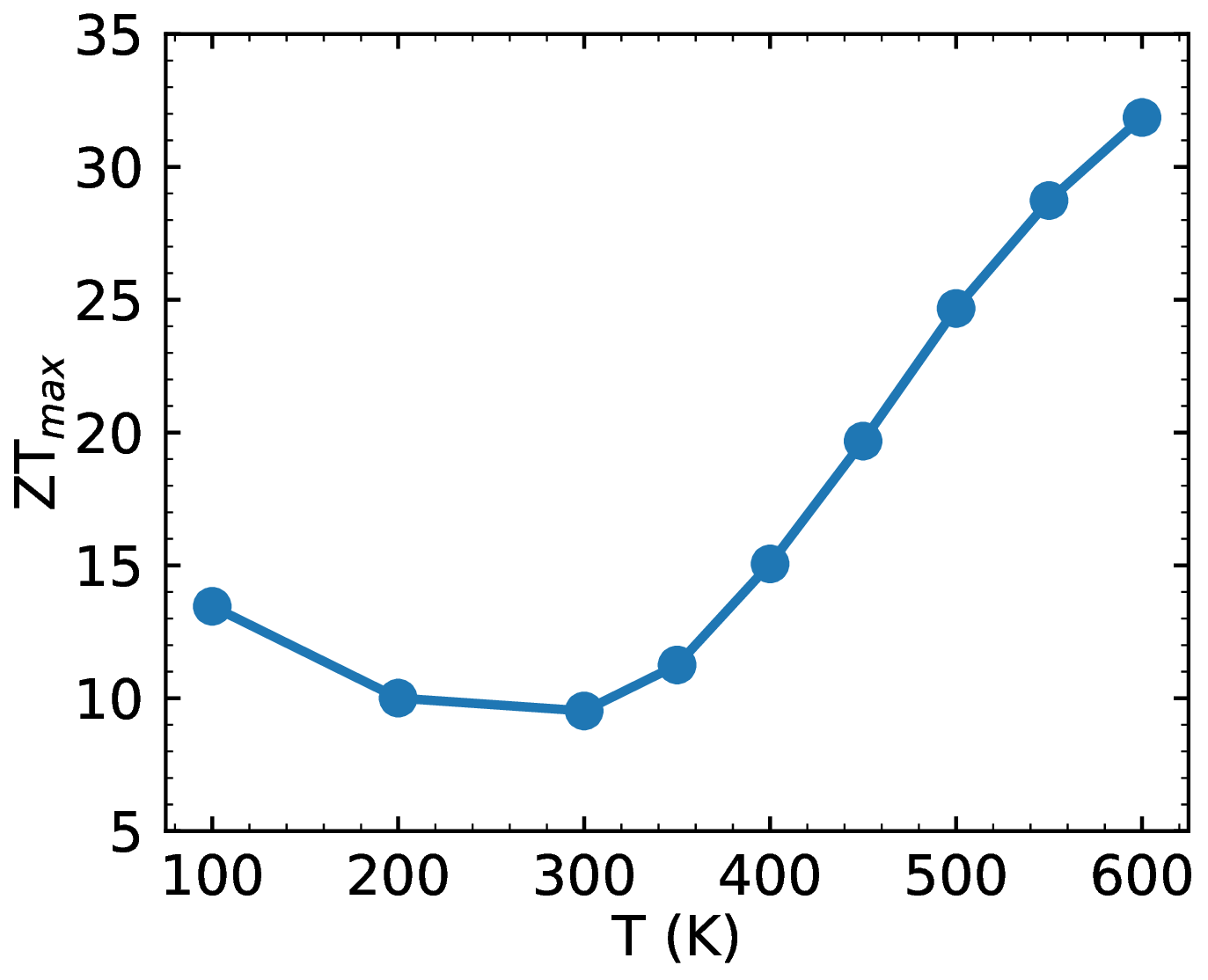}
	%\captionsetup{singlelinecheck=false, justification=justified}
	\caption{$ZT_{\mathrm{max}}$ as a function of temperature T for irradiated Wse$_2$ at L= $30\,$nm and W= $15\,$nm. All other system parameters and the light parameters are the same as in Fig.~\ref{fig:transm}.}
	\label{fig:ZTvT}
\end{figure}
All of the TE quantities discussed so far have been calculated at room temperature, $T = 300\,$K. We now examine how $ZT$ varies with temperature by calculating $ZT_{\text{max}}$ as a function of temperature $T$, as shown in Fig.~\ref{fig:ZTvT}. The temperature is varied from 100 to 600$\,$K and $ZT_{\text{max}}$ is found to increase monotonically with increasing temperature, which indicates its potential for high-temperature thermoelectric applications.

\section{Conclusion}
In this work, we systematically investigate the thermoelectric response of monolayer WSe$_2$ under irradiation by arbitrarily polarized light. The electronic transport coefficients, namely the electrical conductance, Seebeck coefficient, and electronic thermal conductance, are computed within a tight-binding framework that incorporates light matter interaction. The electronic transmission probability is evaluated using the KWANT package within the Landauer-B\"{u}ttiker formalism. The lattice thermal conductance is independently obtained from first principles density functional theory calculations. Combining the electronic and phononic contributions, we determine the thermoelectric figure of merit $ZT$.

Irradiation strongly suppresses the transmission probability, leading to a reduction in both the electrical conductance and the electronic thermal conductance. In contrast, the Seebeck coefficient is significantly enhanced. The simultaneous reduction of thermal transport and enhancement of thermopower results in a marked improvement in thermoelectric efficiency.

We further clarify the role of spin-orbit coupling in shaping the transport properties of monolayer WSe$_2$. The inclusion of spin-orbit coupling induces pronounced band splitting near high symmetry points and modifies the density of states close to the band edges, thereby influencing charge transport while preserving the overall band topology.

On the phonon side, spin-orbit coupling modifies the dispersion by softening low frequency acoustic branches and inducing avoided crossings, which enhance phonon scattering. The resulting increase in anharmonicity, reduction in group velocities, and enlarged scattering phase space significantly suppress the lattice thermal conductivity. The calculated $\kappa_{\text{latt}}$ including spin-orbit coupling shows improved agreement with experimental trends, demonstrating the necessity of incorporating spin-orbit effects for quantitatively reliable transport predictions in WSe$_2$.

Overall, the combined effects of irradiation and spin-orbit coupling produce a robust enhancement of thermoelectric performance, with $ZT$ exceeding unity. The enhancement persists across different system lengths and widths, and over a broad temperature range, indicating the stability of the effect and the potential of light controlled transport engineering in two dimensional transition metal dichalcogenides.

%%%%%%%%%%%%%%%%%%%%%%

%%%%%%%%%%%%%%%%%%%%%%%%%%%%%%%%%%%


\begin{thebibliography}{99} %Produces the bibliography via BibTeX.
\bibitem{mahan1998} G. D. Mahan, {\it Good Thermoelectrics}, Solid State Phys. {\bf 51}, 81 (1998).

\bibitem{disalvo} F. J. DiSalvo, {\it Thermoelectric Cooling and Power Generation}, Science {\bf 285}, 703 (1999).

\bibitem{rowe} D. M. Rowe, ed., {\it CRC Handbook of Thermoelectrics}, (CRC Press, 1995).

\bibitem{goldsmid} H. J. Goldsmid, {\it Introduction to Thermoelectricity}, (Springer, 2010).

\bibitem{bell} L. E. Bell, {\it Cooling, heating, generating power, and recovering waste heat with thermoelectric systems}, Science {\bf 321}, 1457 (2008).

\bibitem{benenti} G. Benenti, G. Casati, and R. S. Whitney, {\it Fundamental aspects of steady-state conversion of heat to work at the nanoscale}, Phys. Reps. {\bf 694}, 1 (2017).

\bibitem{hicks1}L. D. Hicks and M. S. Dresselhaus, \textit{Effect of quantum-well
structures on the thermoelectric figure of merit}, Phys. Rev.
B {\bf 47}, 12727 (1993).

\bibitem{hicks2}L. D. Hicks and M. S. Dresselhaus, \textit{Thermoelectric figure of merit of a one-dimensional conductor}, Phys. Rev. B {\bf 47}, 16631 (1993).

\bibitem{osuala} C. I. Osuala, T. Choudhary, R. K. Biswas, S. Ganguly, and S. K. Maiti, {\it Thermolectricity in irradiated bilayer graphene flakes}, J. Phys. Chem. C {\bf 129}, 3392 (2025).

\bibitem{mahanpnas} G. D. Mahan and J. O. Sofo, {\it The best thermoelectric}, Proc. Natl. Acad. Sci. U.S.A. {\bf 93},  7436 (1996).

\bibitem{heremans} J. P. Heremans,  M. S. Dresselhaus, L. E. Bell, and  D. T. Morelli {\it When thermoelectrics reached the nanoscale}, Nat. Nanotechnol. {\bf 8}, 471 (2013).

\bibitem{etms}  S. Datta, {\it Electronic Transport in Mesoscopic Systems}, Cambridge University Press, Cambridge, 1995.

\bibitem{butler} S. Z. Butler, S. M. Hollen, L. Cao, Y. Cui, J. A. Gupta, H. R. Guti\'{e}rrez, T. F. Heinz, S. S. Hong, J. Huang, A. F. Ismach, et al., {\it Progress, challenges, and opportunities in two-dimensional materials beyond graphene}, ACS nano {\bf 7}, 2898 (2013).

\bibitem{geim} A. K. Geim and I. V. Grigorieva, {\it Van der Waals heterostructures}, Nature {\bf 499}, 419 (2013).

\bibitem{li2018} X. Li and J. Yang, {\it Thermoelectric properties of two-dimensional transition metal dichalcogenides}, Journal of Mat. Chem, A {\bf 6}, 10930 (2018).

\bibitem{ztian2012} Z. Tian, J. Garg, K. Esfarjani, T. Shiga, J. Shiomi, and G. Chen, {\it Phonon conduction in PbSe, PbTe, and PbTe$_{1-x}$Se$_x$ from first-principles calculations}, Phys. Rev. B {\bf 85}, 184303 (2012).

\bibitem{wu2019} T. Wu, X. Chen, H. Xie, Z. Chen, L. Zhang, Z. Pan, and W. Zhuang, {\it Coupling of Spin-Orbit Interaction with Phonon Anharmonicity Leads to Significant Impact on Thermoelectricity in SnSe}, Nano Energy {\bf 60}, 673 (2019).

\bibitem{wli2012} W. Li, L. Lindsay, D. A. Broido, D. A. Stewart, and N. Mingo, {\it Thermal conductivity of bulk and nanowire Mg$_2$Si$_x$Sn$_{1-x}$ alloys from first principles}, Phys. Rev. B {\bf 86}, 174307 (2012).

\bibitem{pyu} P. Yu et al., {\it Metal-Semiconductor Phase-Transition in WSe$_{2(1-x)}$Te$_{2x}$ Monolayer}, Adv. Mater. {\bf 29}, 1603991. (2017).

\bibitem{dhan}	D. Han, X. Yang, M. Du, G. Xin, J. Zhang, X. Wang, and L. Cheng, {\it Improved thermoelectric properties of WS$_2$-WSe$_2$ phononic crystals: Insights from first-principles calculations}, Nanoscale {\bf 13}, 7176 (2021).

\bibitem{oka2009} T. Oka and H. Aoki, {\it Photovoltaic Hall effect in graphene}, Phys. Rev. B {\bf 79}, 081406 (2009).

\bibitem{kitagawa} T. Kitagawa, T. Oka, A. Brataas, L. Fu, and E. Demler, {\it Transport properties of nonequilibrium systems under the application of light}, Phys. Rev. B {\bf 84}, 235108 (2011).

\bibitem{lindner} N. H. Lindner, G. Refael, and V. Galitski, {\it Floquet topological insulator in semiconductor quantum wells}, Nat. Phys. {\bf 7}, 490 (2011).

\bibitem{lopez2015} A. L\'{o}pez, G. Sun, and J. Schliemann, {\it Floquet engineering of long-range p-wave superconductivity}, Phys. Rev. B {\bf 92}, 045303 (2015).

\bibitem{calvo} H. L. Calvo, H. M. Pastawski, S. Roche, and L. E. F. Foa Torres, {\it Tuning laser-induced band gaps in graphene}, Appl. Phys. Lett. {\bf 98}, 232103 (2012).

\bibitem{khoeini} F. Khoeini, K. Shakouri, and F. Peeters, {\it Peculiar half-metallic state in zigzag nanoribbons of MoS$_2$: Spin filtering}, Phys. Rev. B {\bf 94}, 125412 (2016).

\bibitem{silva-guill} J. \'{A}. Silva-Guill\'{e}n, P. San-Jose, and R. Rold\'{a}n, {\it Electronic band structure of transition metal dichalcogenides from ab initio and Slater--Koster tight-binding model}, Appl. Sci. {\bf 6}, 284 (2016).

\bibitem{kosmider}K. Ko\'{s}mider, J. W. Gonz\'{a}lez, and J. Fern\'{a}ndez-Rossier, {\it Large spin splitting in the conduction band of transition metal dichalcogenide monolayers}, Phys. Rev. B {\bf 88}, 245436 (2013).


\bibitem{groth2014kwant} C. W. Groth, M. Wimmer, A. R. Akhmerov, and X. Waintal, \textit {Kwant: a software package for quantum transport}, New J. Phys. {\bf 16}, 063065 (2014).

\bibitem{delplace}  P. Delplace, A. G\'{o}mez-Le\'{o}n, and  G. Platero, \textit{Merging of Dirac points and Floquet topological transitions in ac-driven graphene},  Phys. Rev. B \textbf{88}, 245422 (2013). 

\bibitem{PhysRevLett.110.200403} A. G\'{o}mez-Le\'{o}n and G. Platero, {\it Floquet-Bloch theory and topology in periodically driven lattices}, Phys. Rev. Lett. {\bf 110}, 200403 (2013).

\bibitem{sambe} H. Sambe, \textit{Steady States and Quasienergies of a Quantum-Mechanical System in an Oscillating Field}, Phys. Rev. A {\bf 7}, 2203 (1973).

\bibitem{grifoni1998driven} M. Grifoni and P. H\"{a}nggi, {\it Driven quantum tunneling}, Phys. Rep. {\bf 304}, 229 (1998).

\bibitem{kallol} K. Mondal, S. Ganguly, and S. K. Maiti, {\it Possible route to efficient thermoelectric applications in a driven fractal network}, Sci. Rep. {\bf 11}, 17049 (2021).

\bibitem{GANGULY2021302} S. Ganguly, S. K. Maiti, and S. Sil, {\it Favorable thermoelectric performance in a Rashba spin-orbit coupled ac-driven graphene nanoribbon}, Carbon {\bf 172}, 302 (2021).

\bibitem{harish} U. H. K. Singha, K. Mondal, S. Ganguly, S. K. Maiti, {\it Photo-induced directional transport in extended SSH chains}, Ann. Phys. {\bf 485}, 170317 (2026).

\bibitem{finch2009} C. M. Finch, V. M. Garc\'{i}a-Su\'{a}rez, and C. J. Lambert, {\it Giant thermopower and figure of merit in single-molecule devices}, Phys. Rev. B {\bf 79}, 033405 (2009). 

\bibitem{zerah} E. Zerah-Harush and Y. Dubi, {\it Enhanced Thermoelectric Performance of Hybrid Nanoparticle--Single-Molecule Junctions}, Phys. Rev. Appl. {\bf 3}, 064017 (2015).

\bibitem{dft} P. Giannozzi et al, {\it QUANTUM ESPRESSO: a modular and open-source software project for quantum simulations of materials}, J. Phys.: Condens. Matter {\bf 21}, 395502 (2009).

\bibitem{pbe1} J. P. Perdew, J. A. Chevary, S. H. Vosko, K. A. Jackson, M. R. Pederson, D. J. Singh, and C. Fiolhais, {\it Atoms, molecules, solids, and surfaces: Applications of the generalized gradient approximation for exchange and correlation}, Phys. Rev. B {\bf 46}, 6671 (1992).

\bibitem{pbe2} J. P. Perdew, K. Burke, and M. Ernzerhof, {\it Generalized Gradient Approximation Made Simple}, Phys. Rev. Lett. {\bf 77}, 3865 (1996). 

\bibitem{wang2021} Y. Wang, P. Wisesa, A. Balasubramanian, S. Dwaraknath, and T. Mueller, {\it Rapid generation of optimal generalized Monkhorst-Pack grids}, Comput. Mater. Sci. {\bf 187}, 110100 (2021).

\bibitem{dfpt} S. Baroni, S. de Gironcoli, A. D. Corso, and P. Giannozzi, {\it Phonons and related crystal properties from density-functional perturbation theory}, Rev. Mod. Phys. {\bf 73}, 515 (2001).

\bibitem{bte} W. Li, J. Carrete, N. A. Katcho, N. Mingo, {\it ShengBTE: A Solver of the Boltzmann Transport Equation for Phonons}, Comput. Phys. Commun. {\bf 185}, 1747 (2014). 

\bibitem{ozbal} G. \"{O}zbal, R. T. Senger, C. Sevik, and H. Sevin\c{c}li, {\it Ballistic thermoelectric properties of monolayer semiconducting transition metal dichalcogenides and oxides}, Phys. Rev. B {\bf 100}, 085415 (2019).

\bibitem{easy} E. Easy, Y. Gao, Y. Wang, D. Yan, S. M. Goushehgir, E. H. Yang, B. Xu, and X. Zhang, {\it Experimental and Computational Investigation of Layer-Dependent Thermal Conductivities and Interfacial Thermal Conductance of One- To Three-Layer WSe$_2$}, ACS Appl. Mater. Interfaces {\bf 13}, 13063 (2021).

\bibitem{boccuni} A. Boccuni et al., {\it Unveiling the Role of Spin Currents on the Giant Rashba Splitting in Single-Layer WSe$_2$}, J. Phys. Chem. Lett. {\bf 15}, 7442 (2024).

\bibitem{zulfi} M. Zulfiqar, Y. Zhao, G. Li, Z. C. Li, and J. Ni, {\it Intrinsic Thermal conductivities of monolayer transition metal dichalcogenides MX$_2$ (M = Mo, W; X = S, Se, Te)}, Sci. Rep. {\bf 9}, 4571 (2019).

\bibitem{affandi} Y. Affandi and M. A. Ulil Absor, {\it Electric field-induced anisotropic Rashba splitting in two dimensional tungsten dichalcogenides WX$_2$ (X: S, Se, Te): A first-principles study}, Physica E Low. Dimens. Syst. Nanostruct. {\bf 114}, 113611 (2019).

\bibitem{tang} Y. Tang et al., {\it Simulation of Hubbard model physics in WSe$_2$/WS$_2$ moir\'{e} superlattices}, Nature {\bf 579}, 353 (2020).

\bibitem{amin} B. Amin, T. P. Kaloni, and U. Schwingenschl\"{o}gl, {\it Strain engineering of WS$_2$, WSe$_2$, and WTe$_2$}, RSC Adv. {\bf 4}, 34561 (2014).

\bibitem{kxchen} K. X. Chen, Z. Y. Luo, D. C. Mo, and S. S. Lyu, {\it WSe$_2$ nanoribbons: New high-performance thermoelectric materials}, Phys. Chem. Chem. Phys. {\bf 18}, 16337 (2016).

\bibitem{blume} M. Blume and R.E. Watson, {\it Theory of spin-orbit coupling in atoms I. Derivation of the spin-orbit coupling constant}, Proc. R. Soc. Lond. A. Math. Phys. Sci. {\bf 270} (1340), 127 (1962).

\bibitem{latzke} D. W. Latzke, W. Zhang, A. Suslu, T.-R. Chang, H. Lin, H.-T. Jeng, S. Tongay, J. Wu, A. Bansil, and A. Lanzara, {\it Electronic structure, spin-orbit coupling, and interlayer interaction in bulk MoS$_2$ and WS$_2$}, Phys. Rev. B {\bf 91}, 235202 (2015).

\bibitem{artificial} M. Polini, F. Guinea, M. Lewenstein, H. C. Manoharan, and V. Pellegrini, {\it Artificial honeycomb lattices for electrons, atoms and photons}, Nat. Nanotechnol. {\bf 8}, 625 (2013).

\bibitem{lindsay2010} L. Lindsay, D. A. Broido and N. Mingo, {\it Flexural phonons and thermal transport in graphene}, Phys. Rev. B {\bf 82}, 115427 (2010).

\bibitem{hung2023} N. T. Hung, {\it The role of spin-orbit interaction in low thermal conductivity of Mg$_3$Bi$_2$}, Appl. Phys. Lett. {\bf 123}, 252109 (2023).

\bibitem{mobaraki2019} A. Mobaraki, C. Sevik, H. Yapicioglu, D. \c{C}aklr, and O. G\"{u}lseren, {\it Temperature-dependent phonon spectrum of transition metal dichalcogenides calculated from the spectral energy density: Lattice thermal conductivity as an application}, Phys. Rev. B {\bf 100}, 035402 (2019).

\bibitem{backman} J. Backman, Y. Lee, and M. Luisier, {\it Phonon-limited transport in two-dimensional materials: A unified approach for ab initio mobility and current calculations}, Phys. Rev. Appl. {\bf 21}, 054017 (2024).

\bibitem{xwu2016} X. Wu, V. Varshney, J. Lee, T. Zhang, J. L. Wohlwend, A. K. Roy, and T. Luo, {\it Hydrogenation of Penta-Graphene Leads to Unexpected Large Improvement in Thermal Conductivity}, Nano Lett. {\bf 16}, 3925 (2016).

\bibitem{peng2016} B. Peng, H. Zhang, H. Shao, Y. Xu, G. Ni, R. Zhang, and H. Zhu, {\it Phonon transport properties of two-dimensional group-IV materials from ab initio calculations}, Phys. Rev. B {\bf 94}, 245420 (2016).

\bibitem{lindsay} L. Lindsay, D. A. Broido, and T. L. Reinecke, {\it First-principles determination of ultrahigh thermal conductivity of boron arsenide: A competitor for diamond?}, Phys. Rev. Lett. {\bf 111}, 025901 (2013).




\end{thebibliography}
\end{document}